\documentclass[11pt,a4paper]{article}
\usepackage{url}
\usepackage{jcappub}
\usepackage{amsfonts}
\usepackage{subfigure}
\usepackage{graphicx}

\renewcommand\[{\left[}

\newcommand{\be}{\begin{equation}}
\newcommand{\ee}{\end{equation}}
\newcommand{\bea}{\begin{eqnarray}}
\newcommand{\eea}{\end{eqnarray}}

\newcommand{\beq}{\begin{equation}}
\newcommand{\eeq}{\end{equation}}

\newcommand{\TRH}{T_\mathrm{RH}}
\newcommand{\TD}{T_\mathrm{D}}
\newcommand{\LCDM}{$\Lambda$CDM}
\newcommand{\LCDMA}{$\Lambda$CDM$+m_a$}

\newcommand{\MeV}{\mathrm{MeV}}

\newcommand{\Neff}{N_\mathrm{eff}}
\newcommand{\eV}{\mathrm{eV}}

\newcommand{\sumnu}{\Sigma m_\nu}

\newcommand{\TP}{PlanckTTTEEE}
\newcommand{\TPB}{PlanckTTTEEE+BAO}
\newcommand{\TPL}{PlanckTTTEEE+lensing}

\begin{document}




\title{Thermal axions  with {multi-eV} masses are possible  in low-reheating scenarios }

\author[a,b]{Pierluca Carenza,}
\author[c]{Massimiliano Lattanzi,}
\author[a,b]{Alessandro Mirizzi,}
\author[c,d]{Francesco Forastieri}

\affiliation[a]{Dipartimento Interateneo di Fisica ``Michelangelo Merlin'', Via Amendola 173, 70126 Bari, Italy}
\affiliation[b]{Istituto Nazionale di Fisica Nucleare - Sezione di Bari, Via Orabona 4, 70126 Bari, Italy}
\affiliation[c]{Istituto Nazionale di Fisica Nucleare, Sezione di Ferrara, Via Giuseppe Saragat 1, I-44122 Ferrara, Italy.}
\affiliation[d]{Dipartimento di Fisica e Scienze della Terra, Universit\`a di Ferrara, Via Giuseppe Saragat 1, I-44122 Ferrara, Italy.}

\emailAdd{pierluca.carenza@ba.infn.it,lattanzi@fe.infn.it, alessandro.mirizzi@ba.infn.it, francesco.forastieri@unife.it}

\abstract{
We revise cosmological mass bounds on hadronic axions in low-reheating cosmological scenarios, with a reheating temperature $T_{\rm RH}~\le 100$ MeV, in light of the latest cosmological observations. In this situation, the neutrino decoupling would be unaffected, while the thermal axion relic abundance is suppressed. Moreover, axions are colder in low-reheating temperature scenarios, so that  bounds on their abundance are possibly loosened. As a  consequence of these two facts, cosmological mass limits on axions are relaxed.
Using state-of-the-art cosmological data and characterizing axion-pion  interactions at the leading order in chiral perturbation theory, we find in the standard case an axion mass bound $m_a < 0.26$~eV. However, axions with masses $m_a \simeq 1$~eV, or heavier, would be allowed for reheating temperatures $T_{\rm RH} \lesssim 80$~MeV. Multi-eV axions would be outside the mass sensitivity of current and planned solar axion helioscopes and would demand new experimental approaches to be detected.}
\maketitle

\section{Introduction}

Cosmological observations, including measurements of the cosmic microwave background (CMB) anisotropies and of the distribution of large scale structures (LSS), are powerful tools to constrain the cosmic history of the universe. These precision measurements provide stringent bounds on particle physics models that seek to account for the matter and energy content of the observed Universe. In this context, a notable example is constituted by neutrino masses for which one gets cosmological bounds that are at least one order of magnitude more stringent than those obtained from laboratory searches~\cite{Lesgourgues:2006nd,Lesgourgues:2014zoa,Archidiacono:2013fha,Capozzi:2017ipn,Lattanzi:2017ubx,RoyChoudhury:2019hls}.
The constraining potential of cosmological measurements has been subsequently applied to the case of other low-mass relics~\cite{Hannestad:2003ye}. In this context, a case that has been studied for more than a decade is constituted by axion hot dark matter~\cite{Hannestad:2005df}. 
The analysis of axion mass bounds  has been performed by different groups using an updated set of cosmological 
data~\cite{Melchiorri:2007cd, Hannestad:2007dd,Hannestad:2010yi,Giusarma:2014zza,DiValentino:2015wba,Archidiacono:2013cha,Archidiacono:2015mda,DiValentino:2015zta} for hadronic axions models,  most notably the Kim-Shifman-Vainshtein-Zakharov (KSVZ) axions~\cite{Kim:1979if,Shifman:1979if},
where the QCD axion coupling to Standard Model fermions is negligible, since it vanishes at tree level.
In particular, the latest analysis including the Planck 2018 data and the Baryon Acoustic Oscillations (BAO) measurements allows one to set a bound $m_a < 0.192$~eV at $95$\% C.L~\cite{Giare:2020vzo}. 
It has been also predicted that future cosmological surveys, like EUCLID would improve the sensitivity to axion mass bound, reaching $m_a \sim 0.15$~eV~\cite{Archidiacono:2015mda} which would be nicely complementary with the reach of the future laboratory experiments, like IAXO~\cite{Armengaud:2019uso}.
More recently the cosmological  mass bounds have been considered  also for  models in which axions have a non negligible coupling with the electrons, like the Dine-Fischler-Srednicki-Zhitnisky (DFSZ)  model~\cite{Dine:1981rt,Zhitnitsky:1980tq}.
In this scenario, in~\cite{Ferreira:2020bpb} it has been shown that using the latest Planck and BAO data one would get a bound 
$m_a \lesssim 0.2$~eV when the axion-pion coupling is maximal. However, constraints on $m_a$ may be significantly relaxed and 
possibly vanish if the axion-pion coupling is small. 
Furthermore, we note that future CMB experiments like CMB-S4 will open a new observational window on axions through their sensitivity to the effective number of relativistic degrees of freedom $\Neff$~\cite{Baumann:2016wac,Abazajian:2016yjj,Ferreira:2018vjj}. In the case of the QCD axion, this would allow to probe the region of masses $m_a \sim {\mathcal O}$(few) meV, or even smaller, depending on the particular realization of the model~\cite{Baumann:2016wac}.

The tight cosmological mass bounds are competitive with, and often stronger than, those obtained from stellar energy loss arguments, and from direct laboratory experiments (see~\cite{Irastorza:2018dyq} for a recent review). However, a fundamental questions is to assess  how robust they are with respect to variations of the cosmological model. 
In this context, a common assumption about the history of the Universe is that its expansion was driven by relativistic particles at early epochs. This radiation-dominated era usually arises as a result of the thermalization of the decay products of a massive particle, a process called reheating.
In the standard cosmological model, it is assumed that there was only one such an event, right after primordial inflation, and that it occurred at very large temperatures. However, we cannot \emph{a priori} exclude that there was more than one of such an event, and that the last reheating episode occurred at much lower temperature than those usually associated to inflation.
From a strictly observational point of view, the reheating temperature can be bounded from below using measurements of primordial element abundances and observations of the Cosmic Microwave Background (CMB). Previous analyses have shown the lower bound on the reheating temperature to be ${\mathcal O}$(1 MeV)~\cite{Kawasaki:2000en,Hannestad:2004px,Ichikawa:2005vw,deSalas:2015glj,Hasegawa:2019jsa,Hasegawa:2020ctq}.  Therefore, it is still possible that unstable non-relativistic particles, other than the inflaton, were responsible of more than one reheating processes at different times in the evolution of the Universe, leading to a series of matter and radiation-dominated phases.
In these \emph{low-reheating} scenarios~\cite{Chung:1998rq,Kolb:2003ke,Felder:1999wt}, our Universe could have reheated to a temperature as low as few MeV, postponing the beginning of the radiation-dominated epoch. The cosmological consequences of this model concerning dark matter production and baryogenesis have been widely explored (see, e.g.~\cite{Giudice:2000ex}).
In particular, the consequences on non-thermal cold dark matter axion production and detection has been  studied~\cite{Visinelli:2018wza,Blinov:2019rhb,Blinov:2019jqc,Ramberg:2019dgi}. 
Low-reheating temperature would have also an impact on the decoupling of low-mass thermal relics.
Notably, one can model these low-reheating scenarios through the decay of massive particles $\phi$ with a rate $\Gamma_\phi$, leading to entropy generation. These decays would soften the time evolution of the temperature, increasing the Hubble parameter $H(T)$ and producing an earlier freeze-out of relic particles, suppressing their relic abundance due to entropy generation. 
This effect has been  studied in the context of neutrinos (see, e.g.,~\cite{Kawasaki:2000en,Hannestad:2004px,Ichikawa:2005vw,deSalas:2015glj}), where latest analysis show a relaxation of the mass bound up to $\sum m_\nu <  1$~eV~\cite{deSalas:2015glj}.  A more remarkable  relaxation of the  mass bound was  predicted long time ago for axions by Grin, Smith \& Kamionkowski \cite{Grin:2007yg} (hence thereafter, GSK08).
Motivated by this seminal insight, we perform an updated study of axion mass bounds in low-reheating scenarios, using
the latest cosmological measurements.

The plan of our work is as follows. In Section~2 we compare the axion  thermalization in the standard case and in the low-reheating scenario.
In Section~3 we discuss the cosmological observables in low-reheating scenarios for thermal axions.
In Section~4 we present the cosmological  data we use and we describe our analysis. 
In Section~5 we show and discuss our axion mass bounds in low-reheating scenarios.
Finally, in Section~6 we discuss the phenomenological implications of our findings and we conclude.
An Appendix follows in which we give more details on the  axion decoupling in low-reheating scenario.

\section{Thermal axions}

\subsection{Axion decoupling  in standard cosmology}
 
The most elegant solution to the  {\em strong CP problem} is based on the Peccei-Quinn (PQ) mechanism~\cite{Peccei:1977hh,Peccei:1977ur,Weinberg:1977ma,Wilczek:1977pj}, in which the Standard Model is enlarged with an additional global U(1)$_A$ symmetry, known as the PQ symmetry. 
The axion is the Nambu-Goldstone boson of the PQ symmetry, a low-mass pseudoscalar particle with properties similar to those of neutral pions. 
Recent precision calculations based on chiral perturbation theory~\cite{diCortona:2015ldu} or on lattice QCD~\cite{Borsanyi:2016ksw} predict the axion mass as
\begin{equation}
m_a = \frac{5.7 \,\ \textrm{eV}}{f_a/10^6 \,\ \textrm{GeV}} \,\ ,
\end{equation}
where $f_a$ is the axion decay constant or PQ scale.
The axion interactions with photons, electrons, and hadrons are also controlled by the PQ constant  and scale as  $f_a^{-1}$.
Therefore, the PQ scale determines the axion phenomenology and is constrained by different experiments and astrophysical arguments that involve interactions with photons, electrons, and hadrons (see~\cite{Tanabashi:2018oca,Irastorza:2018dyq,Giannotti:2017hny} for recent reviews). 

Apart from the original theoretical motivation, a renewed interest arose towards axions in the recent years since these still elusive particles can play a crucial role in explaining the dark matter (DM) puzzle in the Universe~\cite{Preskill:1982cy,Dine:1982ah,Abbott:1982af,Sikivie:2006ni}. Indeed, depending on their mass and production mechanism, axions can play the role of both cold and hot dark matter relics. In particular, axions with masses  in the range 10--1500 $\mu$eV might
provide the dominant cold dark matter component, and would be searched by the ADMX~\cite{Du:2018uak,Braine:2019fqb} and MADMAX~\cite{TheMADMAXWorkingGroup:2016hpc} experiments. In this context their main production mechanism would be the non-thermal realignment, and depending on the cosmological scenario there can be some contribution associated with decays of topological defects, like cosmic strings and domain walls.
Furthermore, for a Peccei-Quinn constant $f_a < 1.2 \times 10^{12}$~GeV, there would be also a primordial axion population produced in the hot thermal plasma~\cite{Masso:2002np,Graf:2010tv,Salvio:2013iaa}.
In particular, if axions have masses $m_a \gtrsim 0.15$~eV (i.e. $f_a \lesssim 3.8 \times 10^{7}$~GeV) they would decouple after the QCD phase transition ($T_{\rm QCD} \simeq 200$~MeV), as shown in Fig.~1 of~\cite{Archidiacono:2013cha}. In this case, their most generic interaction processes
would involve pions rather than quarks and gluons present at earlier epochs~\cite{Moroi:1998qs,Chang:1993gm}. These processes would lead to a background of thermal axions, that would provide another hot dark matter component, in addition to active neutrinos.

In principle the region with $f_a < 4.0 \times 10^{8}$~GeV corresponding to $m_a > 15$~meV would be excluded from the observation of SN 1987A neutrinos~\cite{Raffelt:2006cw,Carenza:2019pxu} (see also~\cite{Carenza:2020cis}).
Indeed, an excessive axion emission via nucleon-nucleon bremsstrahlung in the supernova core, would have shortened the observed supernova neutrino burst. However, due to the sparseness of the supernova neutrino data and the uncertainties in the SN modeling at late time, this bound should be taken \emph{cum grano salis}, and it is not redundant to investigate this region of the axion parameter space with other arguments.
In this context, for  models in which axions have a non negligible coupling with the electrons, like the DFSZ model~\cite{Dine:1981rt,Zhitnitsky:1980tq}, there 
are different constraints from different stellar systems (e.g. red giants) which might even exclude masses larger than 10 meV in specific realizations
of the model (see Fig.~12 of Ref.~\cite{DiLuzio:2020wdo}). 
On the other hand, in the case of hadronic axion models, most notably the KSVZ axions~\cite{Kim:1979if,Shifman:1979if}, the strong astrophysical constraints on the axion electron coupling~\cite{Capozzi:2020cbu,Straniero:2020iyi} provide only a weak bound on the PQ constant, since the coupling to electrons is suppressed by loop effects. 
Therefore, one has to rely on the weaker horizontal branch bound~\cite{Ayala:2014pea,Straniero:2015nvc}, $ g_{a\gamma}\lesssim 0.65\times 10^{-10} $~GeV$ ^{-1} $, which translates into $ f_a\geq 1.8 \,C_{\gamma}\times 10^{7}$~GeV (i.e.,  $m_a \le 0.32 \,C_\gamma^{-1}$~eV), where $ C_{\gamma} $ is a model dependent constant.
In this context, models have been proposed where $C_{\gamma}$ is very small, relaxing the axion mass bound~\cite{Kaplan:1985dv}.

In the following we will focus on hadronic axions.
Since these do not couple to charged leptons, their main thermalization process is given by the interactions with pions~\cite{Hannestad:2005df}
\begin{equation}
a + \pi \leftrightarrow \pi+ \pi \,\ , 
\end{equation}
where the axion-pion interaction at the leading order based on effective field theory is described by the  Lagrangian~\cite{DiVecchia:1980yfw,Georgi:1986df,Chang:1993gm,DiLuzio:2021vjd} (see also~\cite{DiLuzio:2020wdo})
\begin{equation}
{\mathcal L}_{a \pi} = \frac{C_{a\pi}}{f_\pi f_a}( \pi^0 \pi^+ \partial_\mu  \pi^- +
\pi^0 \pi^- \partial_\mu  \pi^+ - 2 \pi^+  \pi^- \partial_\mu  \pi^0)\partial^\mu a \,\ ,
\label{eq:lagr}
\end{equation}
where in hadronic axion models, the coupling constant  $C_{a\pi}= (1-z)/[3(1+z)]$, with $z= m_u/m_d=0.48$ is the ratio of up and down quark mass, and $f_\pi = 92$~GeV is the pion decay constant~\cite{diCortona:2015ldu}.
In a recent paper (Ref.~\cite{DiLuzio:2021vjd}) the validity of the calculation of axion thermalization based on the leading order Lagrangian of Eq.~(\ref{eq:lagr}) has been questioned  for decoupling temperatures $T_D \gtrsim 62$~MeV. Indeed, in this case the effective field theory would break down. The authors of Ref.~\cite{DiLuzio:2021vjd}  propose using lattice QCD techniques to avoid a chiral expansion in regimes where its validity may fail. At this stage, however, there is no feasible alternative approach to estimate the thermal axion abundance. Therefore, we limit ourselves to the traditional treatment of axion thermalization  presented in~\cite{Hannestad:2005df}  and comment later on this important point.

Axion decoupling occurs when the interaction rate becomes slow compared with the cosmic expansion rate.
We thus use as criterion for axion decoupling
\begin{equation}
\langle \Gamma_a \rangle = H(T) \,\ ,
\end{equation}
where $\langle \Gamma_a \rangle $ is the axion absorption rate, averaged over a thermal distribution at temperature $T$, whereas $H(T)$ is the Hubble expansion parameter at the cosmic temperature $T$: 
 \begin{equation}
 H(T) = \left[\frac{4 \pi^3}{45} g_\ast(T) \right]^{1/2} \frac{T^2}{m_{\rm Pl}} \,\ .
 \label{eq:hubble}
 \end{equation}
Here, $g_\ast(T)$ is the effective number of thermal degrees of freedom that are excited at the epoch with temperature $T$, and $m_{\rm Pl}=G_N^{-1/2}$ is the Planck mass. Our freeze-out criterion is accurate up to a constant of order unity.
We notice that for decoupling temperatures $T_D > T_{\rm QCD} \gtrsim 200$~MeV, the freeze-out epoch suddenly jumps to a much higher temperature because axion interactions with gluons and quarks before confinement are much less efficient~\cite{Archidiacono:2015mda}.
When needed, we take into account the effect of the QCD phase transition in the effective degrees of freedom
$g_{\ast}$  following the treatment given in~\cite{Husdal:2016haj}, while for simplicity we do not change the interaction rate.
This choice does not affect our results since it involves a region of the parameter space where already the change 
in $g_{\ast}$ is enough to suppress a thermal axion population frozen out before the QCD epoch.

Having determined the axion decoupling temperature $T_{D}$ in this way, one can calculate the present-day axion number density by resorting to entropy conservation. This yields
\begin{equation}
n_a = \frac{g_{\ast S}(T_{\rm today})}{g_{\ast S}(T_{D})} \times \frac{n_\gamma}{2} \,\ ,
\end{equation}
where $g_{\ast S}(T)$ is the effective number of entropy degrees of freedom at temperature $T$, while $T_\mathrm{today} = 2.73$~K and $n_\gamma$ are the present-day temperature and number density of CMB photons.

\subsection{Axion decoupling in low-reheating scenario} 
The axion thermalization described in the previous Section would be strongly affected in nonstandard cosmological  histories, as pointed out in GSK08~\cite{Grin:2007yg}. 
A well-motivated class of nonstandard cosmological histories, is constituted by those realized in low-reheating temperature scenarios. 
Details of axion thermalization in this case are given in the Appendix.

In low-rehating scenarios one considers a  massive particles $\phi$  decaying with a rate $\Gamma_\phi$ into relativistic particles, reheating the Universe in the process. 
The equation for the energy density of $\phi$ is that of a decaying non-relativistic species~\cite{Hannestad:2004px}
\begin{equation}
\frac{d \rho_\phi}{dt} = -\Gamma_\phi \rho_\phi - 3 H \rho_\phi \,\ .
\end{equation}
One can define a reheating temperature $T_{\rm RH}$ as\footnote{Note that in other works, e.g. in~\cite{Hannestad:2004px}, the reheating temperature has been defined
through $\Gamma_\phi= 3 H(T_{\rm RH})$}~\cite{Grin:2007yg} 
\begin{equation}
\Gamma_\phi=  H(T_{\rm RH}) \,\ ,
\label{eq:reheat}
\end{equation}
that marks the point where the Universe is already dominated by radiation with $T_{\rm RH}$. 
From Eqs.~(\ref{eq:hubble}) and~(\ref{eq:reheat}) one obtains
\begin{equation}
T_{\rm RH} \simeq 0.7 \left(\frac{g_\ast}{10.75}\right)^{-1/4} 
\left(\frac{\Gamma_\phi}{\textrm{s}^{-1}}\right) \,\ \textrm{MeV} \,\ , 
\end{equation}
where we fixed $g_\ast$ to its Standard Model value, neglecting the axion degree of freedom. 
This temperature gives a rough idea about the reheating temperature when the Universe enters the standard radiation dominated phase.

The outcome of our calculations do not depend on the choice of the initial time if  $t_{i} \ll t(T_{\rm RH})$, provided that the maximum value of temperature that is reached is significantly larger than the neutrino and axion decoupling temperature. 
The photon temperature $T$ decreases as $t^{-1/4}$ when matter dominates and as  $t^{-1/2}$ when relativistic particles fix the cosmological expansion.
Therefore, the Universe  expands faster during reheating than it would during radiation domination, the Hubble parameter being given by~\cite{Grin:2007yg}
\begin{equation}
H= \sqrt{\frac{5 \pi^3 g_\ast^2(T)}{9 g_{\ast, \rm{RH}}}} \left(\frac{T}{T_{\rm{RH}}}\right)^2 \frac{T^2}{m_{\rm pl}} \,\ ,
\end{equation}
where $g_{\ast, \rm{RH}}$ are the thermal degree of freedom at the reheating temperature.
Due to the faster Universe expansion, the equilibrium condition, $\langle \Gamma_a \rangle \gtrsim H(T)$ is harder to be maintained. 
Particles with decoupling temperatures $T_{\rm D} \gtrsim T_{\rm RH}$ come into chemical equilibrium, but then freeze out before reheating completes. Their abundances are then reduced by entropy production during reheating.
In the specific case of axions, as the reheating temperature is lowered, axions freeze out at higher temperatures due to the higher value of $H$~\cite{Grin:2007yg}.
As the reheating temperature is increased, the $T \sim t^{-1/4}$ epoch becomes increasingly irrelevant and one would recover the standard axion-thermalization.
As done in GSK08~\cite{Grin:2007yg}, in  the following we will limit ourselves to $T_{\rm RH} \gtrsim 10$~MeV.
For these values of the reheating temperature the neutrino decoupling proceeds as in the standard scenario~\cite{deSalas:2015glj}.

\begin{figure}[t!]
	\vspace{0.cm}
	\hspace{1.cm}
	\includegraphics[width=0.8\textwidth]{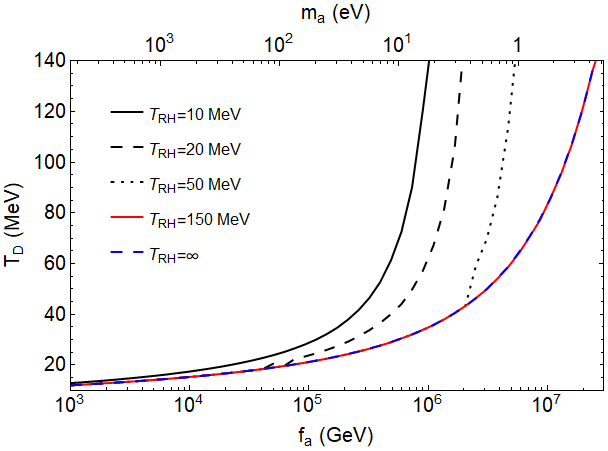}
	\caption{Axion decoupling temperature $T_D$ in function of the Peccei-Quinn scale $f_a$ and the axion mass
	$m_a$ for different reheating temperatures $T_{\rm RH}$.}
	\label{fig:Tdec}
\end{figure}

In Fig.~\ref{fig:Tdec} we show the axion decoupling temperature $T_D$ in function of the Peccei-Quinn scale $f_a$ and the axion mass $m_a$ for different reheating temperatures $T_{\rm RH}$. The standard case corresponds to $T_{\rm RH}= \infty$. We realize that for sufficiently high reheating temperature, $T_{\rm RH}= 150$~MeV in the figure, the behaviour is identical to the standard case. Lowering the reheating temperature we find that for $T_D < T_{\rm RH}$ the decoupling temperature coincides with the one obtained in the standard scenario. Conversely for $T_D > T_{\rm RH}$ one finds a higher decoupling temperature with respect to the standard case due to the higher Universe expansion rate.
	
From the decoupling temperature it is possible to  calculate the present axion abundance as~\cite{Grin:2007yg} 
\beq
\Omega_{a} h^{2}= \frac{m_{a}}{13 \,\ \textrm {eV}}\frac{1}{g_{*,s}(T_{D})}\times \Bigg\{
\begin{array}{cl}
\left(\frac{T_{\rm RH}}{T_{D}}\right)^{5} \left[\frac{g_{\ast}(T_{\rm RH})}{g_{\ast}(T_{D})}\right]^{2}\left[\frac{g_{\ast,S}(T_{D})}{g_{\ast,S}(T_{\rm RH})}\right] 
& \textrm{for} \,\ T_{D}>T_{\rm RH} \,\ 
\\
1 & \textrm{for} \,\ T_{D}\leq T_{\rm RH} \,\ \\
\end{array}
\eeq

\begin{figure}[t!]
	\vspace{0.cm}
	\hspace{1.cm}
	\includegraphics[width=0.8\textwidth]{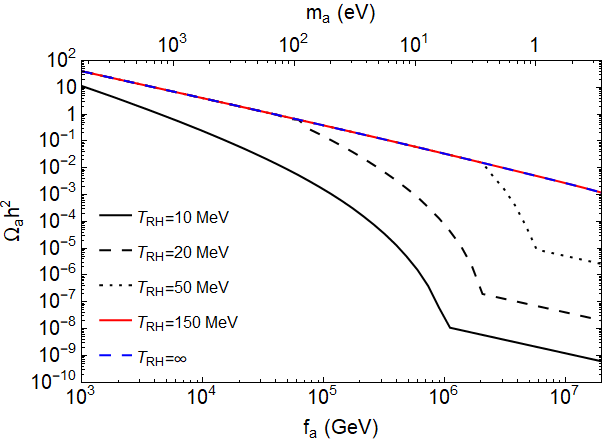}
	\caption{Present axion abundance  $\Omega_{a} h^{2}$  in function of the Peccei-Quinn scale $f_a$ and the axion mass
	$m_a$ for different reheating temperatures $T_{\rm RH}$.}
	\label{fig:abund}
\end{figure}

In Fig.~\ref{fig:abund} we plot the axion abundance $\Omega_{a} h^{2}$ in function of the Peccei-Quinn scale $f_a$ and the axion mass $m_a$ for different reheating temperatures $T_{\rm RH}$. One realizes that lowering the reheating temperature the axion abundance gets significantly suppressed with respect to the standard case. In particular, the lower the reheating temperature $T_{\rm RH}$ the higher the axion mass for which one starts finding the suppression in the abundance.
We also realize that for $T_{\rm RH} < T_{\rm QCD }$~MeV there is a  change in the slope in $\Omega_{a} h^{2}$ that becomes
milder at larger $f_a$. This change corresponds to axion decoupling occurring after the QCD phase transition.
	
\begin{figure}[t!]
	\vspace{0.cm}
	\hspace{1.cm}
	\includegraphics[width=0.8\textwidth]{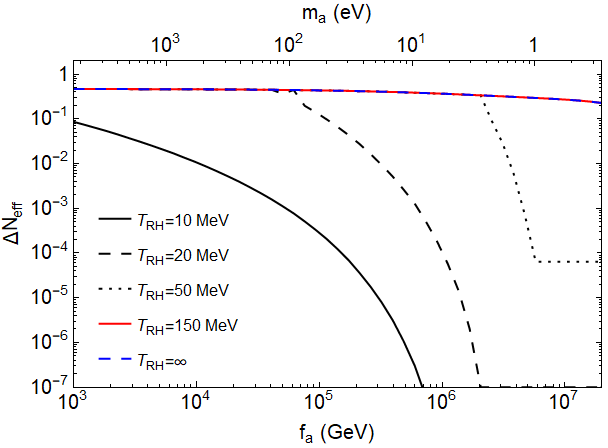}
	\caption{Axion extra-radiation $\Delta N_{\rm eff}$ at $T=1$~keV in function of the Peccei-Quinn scale $f_a$ and the axion mass $m_a$ for different reheating temperatures $T_{\rm RH}$.}
	\label{DNeff}
\end{figure}

Axions may also give contribution to extra-radiation, affecting the effective number of relativistic species $N_{\rm eff}$. As long as $T_a \gg m_a$, this extra-contribution can be evaluated as~\cite{Grin:2007yg}  
\beq
\Delta N_{\rm eff}= \left(\frac{4}{7}\right)\left(\frac{11}{4}\right)^{4/3} \times \Bigg\{
\begin{array}{ll}
\left(\frac{T_{\rm RH}}{T_{D}}\right)^{20/3} \left[\frac{g_{\ast}(T_{\rm RH})}{g_{\ast}(T_{D})}\right]^{8/3}\left[\frac{g_{\ast,S}(T)}{g_{\ast,S}(T_{\rm RH})}\right]^{4/3} &
\textrm{for} \,\ T_{D}>T_{\rm RH}  \\
\left[\frac{g_{\ast,S}(T)}{g_{\ast,S}(T_{D})}\right]^{4/3} & \textrm{for} \,\ T_{D}\leq T_{\rm RH} \,\ \\
\end{array}
\eeq
In Fig.~\ref{DNeff} we show the axion extra-radiation $\Delta N_{\rm eff}$ in function of the Peccei-Quinn scale $f_a$ and the axion mass $m_a$ for different reheating temperatures $T_{\rm RH}$. In the standard case an axion with $m_a \sim 1$ eV would contribute with $\Delta N_{\rm eff}\simeq 0.29$. Once more we realize that lowering the reheating temperature one finds a dramatic suppression of $\Delta N_{\rm eff}$ with respect to the standard case.

\section{Thermal axions and cosmological observables in low-reheating scenarios \label{sec:axions_cosmo}}

We now build on the results presented in the previous sections, and describe the imprint of thermal axions in low-reheating scenarios on cosmological observables.
In the standard scenario, thermal axions behave as hot dark matter, i.e., their average momentum $\langle p_a \rangle \sim T_a \gg m_a$ at the times of interest for structure formation. Their phenomenology is related to their large free-streaming length, and their imprint on cosmological observables is very similar to that of the active neutrinos.
This is however not necessarily the case in low-reheating scenarios, as also noted in GSK08~\cite{Grin:2007yg}. In fact, when $\TRH \ll \TD$, the entropy generation occurring during the reheating stage can strongly suppress the axion-to-photon temperature ratio, effectively resulting in a much lower axion temperature with respect to the standard scenario, as we show in the following.

\begin{figure}[t!]
	\vspace{0.cm}
	\hspace{1.cm}
	\includegraphics[width=0.9\textwidth]{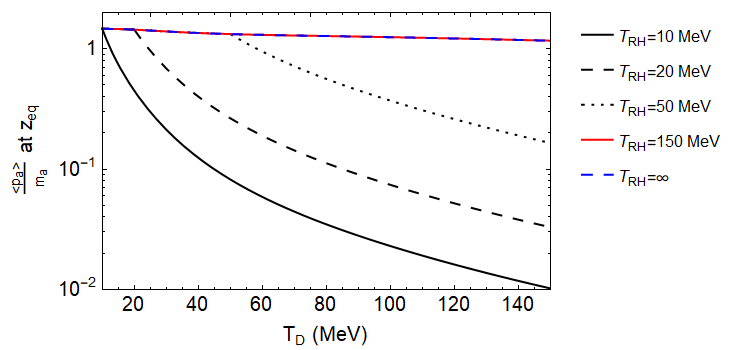}
	\caption{Axion average momentum-to-mass ratio $\langle p_a\rangle/m_a$  at redshift $z_{\rm eq}=3400$ and for an axion mass $m_{a}=1$~eV in function of the decoupling temperature for different reheating temperatures $T_{\rm RH}$.}
	\label{fig:poverm}
\end{figure}

The axion-to-photon temperature ratio ${\mathcal T}_a$ well after reheating and decoupling ($T \ll \TRH, \TD$) is given by
\begin{equation}
{\mathcal T}_a \equiv \frac{T_a}{T} = 
\left\{
\begin{array}{ll}
\displaystyle \left(\frac{\TRH}{\TD}\right)^{5/3} \left[\frac{g_*(\TRH)}{g_*(\TD)}\right]^{2/3} \left[\frac{g_{*S}(T)}{g_{*S}(\TRH)}\right]^{1/3} & \quad \textrm{for } \TD\gg \TRH\,, \\[0.5cm]
\displaystyle \left[\frac{g_{*S}(T)}{g_{*S}(\TD)}\right]^{1/3} & \quad \textrm{for } \TD\ll \TRH \, .\\
\end{array}
\right.
\end{equation}
From this, the thermally-averaged ratio of axion momentum to mass at a given redshift $z$ can be readily computed as
\begin{equation}
\frac{\langle p_a\rangle}{m_a} =  2.7\,\frac{T_\mathrm{today} (1+z)}{m_a}\times
\left\{
\begin{array}{ll}
\displaystyle   \left(\frac{\TRH}{\TD}\right)^{5/3} \left[\frac{g_*(\TRH)}{g_*(\TD)}\right]^{2/3} \left[\frac{g_*(T_\mathrm{today})}{g_*(\TRH)}\right]^{1/3} & \quad \textrm{for } \TD\gg \TRH\,, \\[0.5cm]
\displaystyle \left[\frac{g_*(T_\mathrm{today})}{g_*(\TD)}\right]^{1/3} & \quad \textrm{for } \TD\ll \TRH \, ,\\
\end{array}
\right.
\end{equation}
where we have used $\langle p_a \rangle = 2.7~T_a$ for bosons, $T \propto g_{*s}^{1/3} (1+z)$. It is clear from the above expression that the ratio $\langle p_a\rangle/m_a$ can be much smaller with respect to the standard case when $\TD \gg \TRH$. This behavior is shown in Fig.~\ref{fig:poverm} for an axion mass  $m_{a}=1$~eV, at the redshift of matter-radiation equality $z_{\rm eq}\simeq 3400$.

The cosmological phenomenology of axions, and of relic particles in general, crucially depends on the value of $\langle p_a\rangle/m_a$ at the times probed by observations. In the standard case $\langle p_a\rangle/m_a \gg 1$ for a large fraction of the cosmic history for thermal axions, and these behave as \emph{hot dark matter}. In this scenario the axion density is severely constrained both by CMB measurements of $\Neff$, and by their potential effect on structure formation, requiring that only a subdominant component of the total matter density can be contributed by particles with a large velocity dispersion. In low-reheating scenarios, however, the condition $\langle p_a\rangle/m_a \ll 1$ might be realized even at early times, and thermal axions would effectively behave as \emph{cold dark matter}. In this scenario, axions would not contribute to $\Neff$, and bounds on their density would come only from the (weaker) requirement that this should not exceed the cold dark matter density inferred e.g. through measurements of CMB anisotropies. 

A good proxy to assess the ``hot'' or ``cold'' nature of a relic is its behaviour at the time of matter-radiation equality\footnote{Note that, in principle, the redshift of matter-radiation equality depends on the cosmological parameters and in particular on $m_a$. However, for the purpose of the qualitative considerations made in this section, we fix this quantity to its  $\Lambda$CDM estimate from Planck 2018~\cite{Aghanim:2018eyx}.}
$z_\mathrm{eq}\simeq 3400$, when the growth of density perturbations effectively starts.~\footnote{During the radiation era, density perturbations inside the horizon can grow logarithmically, at most (Meszaros effect). }
For example, in their classic textbook Kolb \& Turner~\cite{Kolb:1990vq} compare the free-streaming length of the relic at $z_\mathrm{eq}$ to the Hubble radius at the same time, as a criterion to discriminate hot and cold dark matter. Similarly, one could evaluate the average momentum-to-mass ratio at $z_\mathrm{eq}$.

\begin{figure}[t!]
	\vspace{0.cm}
	\includegraphics[width=0.45\textwidth]{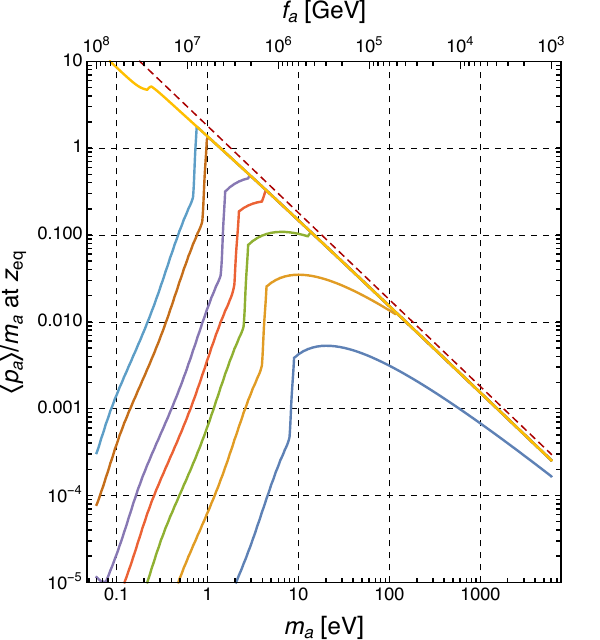}
	\includegraphics[width=0.455\textwidth]{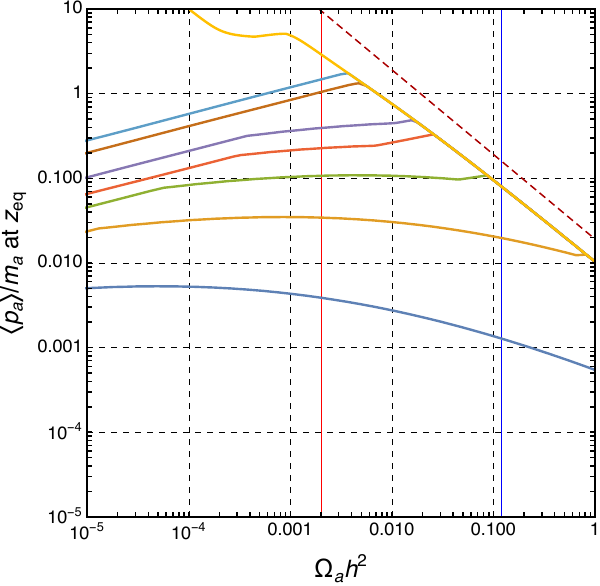}
	\caption{Left panel: Axion momentum-to-mass ratio at redshift $z_\mathrm{eq}=3400$, as a function of the Peccei-Quinn scale $f_a$ and the axion mass
	$m_a$ for different reheating temperatures $T_{\rm RH}$. From top to bottom: $\TRH = (\infty, 100, 80, 50, 40, 30, 20, 10) \,\mathrm{MeV}$. For comparison, the same quantity for active neutrinos is also shown (uppermost dashed curve). Right panel: same as left panel, but as a function of $\Omega_a h^2$. The vertical red and blue lines correspond to the reference values $\Omega_a h^2 = 2 \times 10^{-3}$ and $\Omega_a h^2 = 0.12$, roughly marking the upper limits on the density of a hot ($\langle p_a\rangle/m_a \gg 1$),  or cold ($\langle p_a\rangle/m_a \ll 1$) axion, respectively (see text for discussion).}
	\label{fig:pmratio}
\end{figure}
We show in Fig.~\ref{fig:pmratio}  the $\langle p_a\rangle/m_a$ ratio evaluated at $z=3400$ as a function of the axion mass $m_a$ (left panel) and energy density $\Omega_a h^2$ (right panel), for different values of the reheating temperature. Let us first examine the two extreme cases, i.e. the standard thermal history with $\TRH = \infty$ and the very low reheating scenario with $\TRH= 10\,\mathrm{MeV}$. In the former case, the average momentum-to-mass ratio decreases with axion mass, with axions lighter than $\sim 1.4$~eV having $\langle p_a\rangle/m_a > 1$ at $z_\mathrm{eq}$. In terms of axion density, a mass of $1.4$~eV corresponds to $\Omega_a h^2 \simeq 7\times 10^{-3}$ for $\TRH=\infty$.  Let us compare this value to observational limits on the density of hot relics. As a rough guide, we consider the upper bound on the sum of neutrino masses $\sum m_\nu < 0.257$~eV (at 95\% CL) from Planck 2018 temperature and polarization data~\cite{Aghanim:2018eyx}, and recast it as a bound on the present density parameter $\Omega_h$ of hot relics other than active neutrinos. After subtracting the contribution of active neutrinos with the minimum mass allowed by flavour oscillation experiments $\sum m_\nu = 0.06$~eV~\cite{Capozzi:2017ipn}, we obtain $\Omega_a h^2 \le\Omega_h h^2 \lesssim 2\times 10^{-3}$, corresponding to $m_a \lesssim 0.45$~eV and $\langle p_a \rangle/m_a (z=z_\mathrm{eq}) \simeq 3$. This limit should apply to hot axions, which is certainly the case for $m_a < 1.4$~eV. Hot axions should also satisfy observational constraints on $\Neff$. Imposing $\Delta\Neff < 0.35$, roughly corresponding to the 95\% credible upper limit from the Planck 2018 data~\cite{Aghanim:2018eyx}, yields $m_a < 0.96\,\eV$. This is of the same magnitude, albeit larger, than the limit coming from the present density of hot dark matter, so we can expect that combining the two requirements will yield an upper limit on $m_a$ somehow smaller than $0.45\,\eV$.
Let us now consider the mass range $m_a > 1.4$~eV, where $\langle p_a\rangle/m_a < 1$. Increasing the mass, a regime will be eventually reached where $\langle p_a\rangle/m_a \ll 1$. Here, thermal axions are cold and the relevant bound comes from requirement that their density should not overshoot the value of the cold dark matter density measured by Planck, $\Omega_a h^2 \le \Omega_{c} h^2 = 0.1202 \pm 0.0014$. Since this is a looser requirement than $\Omega_a h^2 \lesssim 2\times 10^{-3}$, in principle it might happen that another allowed region of axion masses exists for $m_a > 1.4$~eV. A closer inspection shows that this is probably not the case for $\TRH = \infty$.
In fact, the value $\Omega_a h^2 = 0.12$ is reached for $m_a \simeq 20$~eV and $\langle p_a\rangle/m_a \simeq 0.1$. Such a value of $\langle p_a\rangle/m_a$ indicates that axions are still ``warm'' at equality (i.e. that their free-streaming length is a sizeable fraction of the Hubble radius). To reach the region in which thermal axions behave as cold, we should go to values of $\Omega_a h^2>0.12$, at variance with Planck observations. Let us now turn our attention to the scenario with a reheating temperature $\TRH= 10\,\mathrm{MeV}$. In this case, the average momentum-to-mass ratio at $z=3400$ is at most $\lesssim 5\times10^{-3}$. We can expect that in this case the thermal axion phenomenology is essentially that of a cold particle, and thus the only relevant constraint is $\Omega_a h^2 \lesssim 0.12$, or $m_a \lesssim 410$~eV.

Based on these considerations, we can draw the following general picture, that should apply also to the scenarios in between the two that have been analysed above. At fixed reheating temperature, the axion mass uniquely fixes the values of $\Omega_a h^2$, $\Delta \Neff$ and $\langle p_a\rangle/m_a$. We expect the likelihood to be a decreasing (or at least, non-increasing) function of all these quantities. Note also that while $\Omega_a h^2$ and $\Delta \Neff$ grow monotonically with $m_a$, $\langle p_a\rangle/m_a$ does not necessarily share this feature, as it can be seen from Fig.~\ref{fig:pmratio}. Thus, in principle it might be possible that, in addition to the expected minimum in $m_a\simeq 0$, the likelihood also has a local minimum at $m_a \ne 0$ and that two disconnected allowed regions for $m_a$ exist. We will see that at the end this is not realized in practice, nevertheless this has been an important point to keep in mind when performing Monte Carlo runs for parameter estimates.

Let us conclude by stressing that the above considerations are qualitative, as it should be clear that the transition from the hot to cold behaviour is a continuous one. Thus we might found that in some part of the parameter space the relevant limits on the axion density parameter $\Omega_a h^2$ and contribution to $\Neff$ lie somewhere in between those for a hot and cold species.  It is also not immediate to estimate the values of $\langle p_a\rangle/m_a (z=z_\mathrm{eq})$ that are, observations-wise, indistinguishable from the limiting cases $\langle p_a\rangle/m_a = 0$ (cold) and $\langle p_a\rangle/m_a = \infty$ (hot). However, this is not a problem given the qualitative nature of the considerations made in this section, whose purpose is only to understand what to expect from the limits that will be derived in the next section through a Boltzmann code that precisely tracks  the evolution of cosmological perturbation.

\section{Data and analysis}

We use the most up-to-date available observations of CMB temperature and polarization anisotropies, possibly in combination with baryon acoustic oscillation (BAO) measurements from galaxy surveys, in order to derive bounds on axion masses in low-reheating scenarios. In particular, we consider CMB observations from the Planck 2018 legacy data release\footnote{Data available at this url:~\url{http://pla.esac.esa.int/pla}.} \cite{Akrami:2018vks} together with BAO measurements from BOSS Data Release 12 \cite{Alam:2016hwk}, 6dFGS \cite{Beutler:2011hx} and SDSS-MGS \cite{Ross:2014qpa}. Our baseline dataset consists of the combination of the Planck \texttt{Plik} likelihood using TT, EE and TE power spectra at multipoles $\ell \ge 30$, and of the \texttt{Commander} and \texttt{SimAll} likelihoods for temperature and EE polarization, respectively, at low-$\ell$'s ($\ell <30$) \cite{Aghanim:2019ame}. This combination is labeled
Planck TT,TE,EE+lowE in the Planck collaboration papers; however, for the sake of brevity we will denote it simply as ``\TP''. However, here and in the following, use of the low-$\ell$ polarization data should be always understood. In fact, large-scale polarization data are needed in order to constrain the optical depth to reionization, $\tau$; this is particularly important for our analysis, since $\tau$ is degenerate with the abundance of hot relics.

In addition to the baseline, we consider two other datasets. In the first, the baseline is augmented by the Planck lensing likelihood \cite{Aghanim:2018oex}: we call this combination ``\TPL''. In the second, labeled ``\TPB'' we add to \TP\ the BAO data described above.

We use a modified version of the Boltzmann code \texttt{camb} to compute theoretical predictions for a given cosmological model.
It is well known that the public version of \texttt{camb} allows for the inclusion of the effects of massive neutrinos in the evolution of cosmological perturbations. Through a suitable remapping of the neutrino parameters, this feature can in fact be extended to include hot or warm relics with a distribution proportional to a Fermi-Dirac, in practice treating them as ``effective neutrinos''  \cite{Colombi:1995ze}. This property has been used, for example, in analyses aiming to derive cosmological bounds on sterile neutrinos, including the one reported in the Planck collaboration papers. However, an exact treatment of bosonic relics is not a feature of the public \texttt{camb} version, because the distribution function of hot/warm relics cannot be changed by the user. One can still resort to the ``effective neutrino'' approach, that however is only approximate, instead than being exact as in the case of fermions. For this work, instead, we have modified \texttt{camb} so that the code correctly uses a Bose-Einstein distribution for axions. In this way, once the axion parameters (essentially the present axion density and their contribution to $\Neff$) are specified, all other quantities (firstly mass and temperature, and secondly the various integrals over the distribution function) are computed consistently.

We consider a family of one-parameter extensions to the standard \LCDM\ cosmological model, labeled \LCDMA\ in which we allow for the presence of thermal axions. In each of these extensions the reheating temperature is fixed; in particular, we consider the values $\TRH= \left\{10,\,20,\,30,\,40,\,50,\,80,\,100 \right\}\,\MeV$, and the case of a standard thermal history, as above refered as $\TRH=\infty$. These models could in principle be described by extending the usual \LCDM\ parameterization with the axion mass,
thus in terms of seven cosmological parameters: the baryon density $\omega_b \equiv \Omega_b h^2$, the cold dark matter (CDM) density $\omega_c \equiv \Omega_c h^2$, the angle subtended by the sound horizon at recombination $\theta_s$, the optical depth to reionization $\tau$, the logarithmic amplitude $\log(10^{10} A_s)$ and the spectral index $n_s$ of the spectrum of primordial scalar perturbations. However, we find that a better parametrization can be obtained by trading the CDM density for the total CDM+axions density $\omega_{c+a} \equiv \omega_c + \omega_a$. The reason is that, for some values of the reheating temperature considered here, axions behave as cold particles, as explained in the previous section\footnote{Note that we always use ``CDM'' to indicate the ``non-axionic'' component of DM.}. In this case $\omega_c$ and $\omega_a$ are strongly degenerate, because only their sum, the total abundance of cold particles, is actually constrained by the data. Thus we prefer to use $\omega_{c+a}$ as a base parameter, and derive the CDM abundance from $\omega_c = \omega_{c+a} - \omega_a$ (the latter being itself a derived parameter computed from $m_a$). In conclusion, this is the vector of base parameters, that take implicit flat priors in our Monte Carlo runs: $\left\{ \omega_b,\,\omega_{c+a},\,\theta_s,\,\tau,\,\log(10^{10} A_s),\,n_s,\,m_a \right\}$. We consider one massive and two massless active neutrino species, with a fixed sum of neutrino masses $\sumnu = 0.06\,\eV$. The three active neutrino species contribute the standard value $\Neff= 3.046$ to the energy density of relativistic species after Big Bang Nucleosynthesis (BBN)\footnote{Refined calculations \cite{Akita:2020szl,Bennett:2020zkv} have recently modified this value to $\Neff=3.044$, but the difference is not relevant for the datasets considered in this work.}. We assume flatness and adiabiatic initial conditions. 

We use the Markov Chain Monte Carlo engine \texttt{CosmoMC}, interfaced with our modified version of \texttt{camb} and with the likelihood modules described above, to explore posteriors and derive estimates and uncertainties for the parameters of the model, and in particular to obtain bounds on the axion mass in low-reheating scenario. In addition to the base parameters, we also include nuisance parameters required by the likelihoods, that are eventually marginalized over. We check convergence of the chains by monitoring the Gelman-Rubin $R-1$ parameter. We also check stability of the limits on $m_a$ obtained by considering each chain of a given run, one at a time.~\footnote{We usually generate 6 chains for each parameter estimation run.} Moreover, since we might expect that the distribution is multimodal, for each model/dataset combination we perform (at least) two twin runs, one with Monte Carlo temperature~\footnote{Not to be confused with the reheating temperature.} $T_\mathrm{MC} = 1$ and the other with $T_\mathrm{MC} > 1$ (we usually take $T_\mathrm{MC} = 2$ or $3$). Setting the Monte Carlo temperature to a value other than 1 amounts to sample from ${\mathcal P}^{1/T}$ instead than from $\mathcal P$, with the latter being the posterior probability distribution of the parameters. Once the chains have been generated, they are importance weighted (``cooled'') to provide samples from the posterior $\mathcal P$. This procedure makes it easier to sample multimodal distributions, at the expense of a slower convergence. We use the ``high-$T_\mathrm{MC}$'' runs to check for the existence of separate peaks in the probability distribution (possibly corresponding to hot and cold axions). We also check consistency of the limits obtained between each pair of twin runs. 

\section{Results and discussions}

\begin{table}[tbp]
\centering
\begin{tabular}{|l|c|c|c|}
\hline
 & \TP & \TPL & \TPB \\
\hline
$\TRH = 10\,\MeV$& $< 391$ & $< 393$  &$ < 391$\\
$\TRH = 20\,\MeV$& $< 39.2$ & $< 36.6$  &$ < 36.0$\\
$\TRH = 30\,\MeV$& $< 6.72$ & $< 6.55$  &$ < 6.31$\\
$\TRH = 40\,\MeV$& $< 3.23$ & $< 3.22$  &$ < 3.05$\\
$\TRH = 50\,\MeV$& $< 2.09$ & $< 2.05$  &$ < 1.89$\\
$\TRH = 80\,\MeV$& $< 0.912$ & $< 0.906$  &$ < 0.901$\\
$\TRH = 100\,\MeV$& $< 0.691$ & $< 0.696$  &$ < 0.688$ \\
$\TRH = \infty $& $<0.837$ & $<0.639$ & $<0.259$ \\
\hline
\end{tabular}
\caption{\label{tab:ma_bounds} 95\% CL upper bounds on axion mass $m_a$ (in eV) from different datasets, for the values of the reheating temperature $\TRH$ indicated in the first column.
} 
\end{table}

\begin{figure}[t!]
	\vspace{0.cm}
	\includegraphics[width=\textwidth]{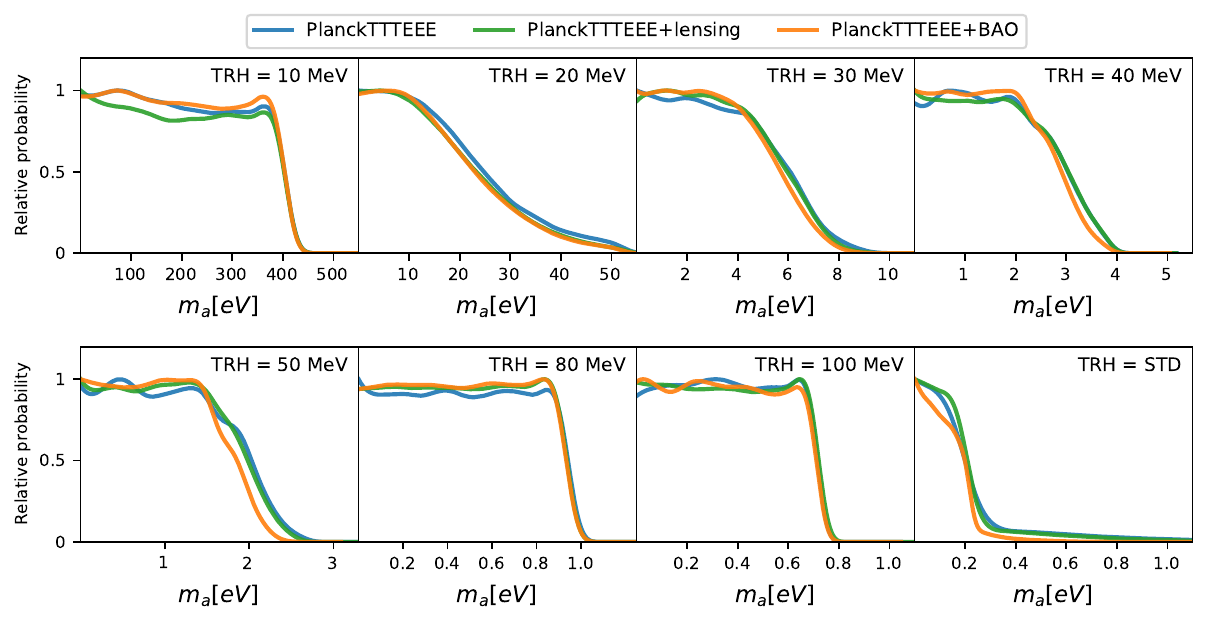}
	\caption{Posterior probability distributions for $m_a$  for different values of the reheating temperature $\TRH$. We show constraints from PlanckTTTEEE alone (blue) and in combination with Planck lensing (green) or BAO data (orange). Each posterior is normalized to its maximum and has been smoothed using a Gaussian kernel with width equal to 0.2 standard deviations of the distribution.}
	\label{fig:post1D_ma}
\end{figure}

Upper limits on the axion mass in low reheating scenarios are reported in Table~\ref{tab:ma_bounds} for different values of the reheating temperature, for the dataset combinations described in the previous section. The corresponding 1-dimensional posterior probability distributions are shown in Fig.~\ref{fig:post1D_ma}.

\begin{figure}[t!]
\begin{center}
	\vspace{0.cm}
	\includegraphics[width=0.48\textwidth]{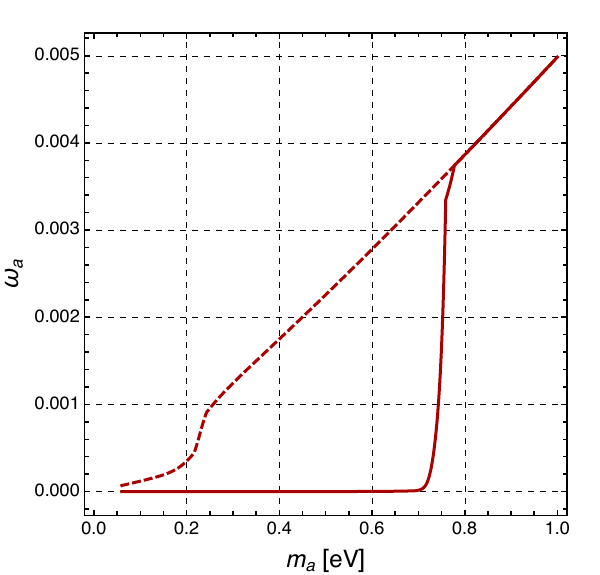}
	\includegraphics[width=0.45\textwidth]{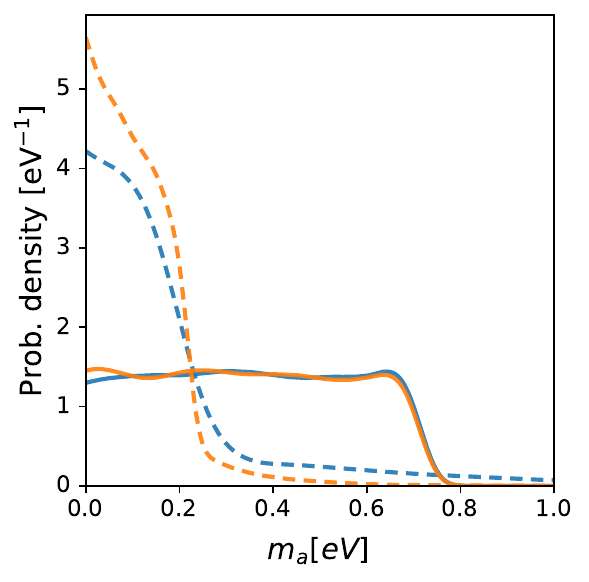}
	\caption{Left panel: axion density parameter $\omega_a$ as a function of the axion mass, for $\TRH = 100$\,MeV (solid) and for the standard thermal history (dashed). Right panel: Posterior probability distributions for $\TRH = 100$\,MeV (solid) and for the standard thermal history (dashed). We show constraints from PlanckTTTEEE alone (blue) and in combination with BAO data (orange). Contrarily to Fig.~\ref{fig:post1D_ma}, here the posteriors are normalized so that they integrate to unity, to better appreciate the difference in the probability densities.}
	\label{fig:oma_vs_ma}
\end{center}
\end{figure}

We start by discussing the constraints obtained from the \TP\ dataset. In the most extreme scenario that we consider, $\TRH = 10\,\MeV$, the upper bound~\footnote{Unless otherwise stated, the upper bounds discussed in the following always correspond to 95\% Bayesian credible intervals.} on the axion mass is relaxed to $m_a < 391\,\eV$. 
As anticipated in the previous sections, this is due to fact that the axion energy density and velocity dispersion are both suppressed with respect to the standard scenario (see Figs.~\ref{fig:abund} and \ref{fig:pmratio}). 
Thus, axions are less abundant for a given mass and lower reheating temperatures; on top of that, they are also colder, so that observational limits on the abundance are relaxed.
In fact, $m_a = 391\,\eV$ and $\TRH = 10\,\MeV$ yield $\omega_a = 0.108$, i.e. axions making up for nearly all the DM content of the Universe. 
This is in agreement with the low average momentum-to-mass ratio (always smaller than $10^{-2}$ at the time of equality) shown in Fig.~\ref{fig:pmratio}. 
For a larger value of the reheating temperature $\TRH = 20\,\MeV$, we find the upper bound $m_a < 39\,\eV$, corresponding to $\omega_a = 0.057$. In this case axions can make up for a large fraction of the DM, but not all of it. We relate this finding to the larger velocity dispersion (i.e., axions are warmer) with respect to the $\TRH = 10\,\MeV$ scenario. 
This trend continues as the reheating temperature increases, and axions of a given mass became warmer and more abundant. The upper bound in the $\TRH = 50\,\MeV$ scenario is $m_a<2.1\,\eV$, corresponding to $\omega_a = 2.4 \times 10^{-3}$. 
The latter value is of the same order of magnitude as the maximum energy density in active neutrinos allowed by Planck data. 
This is again supported by a visual inspection of Fig.~\ref{fig:pmratio}, from which the relatively large ratio at equality $\langle p_a \rangle /m_a \simeq 0.4$, can be read for $m_a \simeq 2 \,\eV$ and $\TRH > 50 \,\MeV$; this is just a factor of 2 smaller than the corresponding value for active neutrinos in the standard scenario (blue curve in the same figure). 
Further increasing the reheating temperature yields $m_a < 0.91\,\eV$ ($\TRH = 80\,\MeV$) and $m_a < 0.69 \,\eV$ ($\TRH = 100\,\MeV$).
For these values of the reheating temperature, the axion density is varying very quickly with $m_a$ when the latter is close to the quoted upper limits, so it is not instructive to express these numbers in terms of $\omega_a$. Finally, we obtain $m_a < 0.84\,\eV$ for the standard thermal history. This bound is actually looser than the one found for $\TRH = 100 \,\MeV$.
This seemingly counterintuitive result can be understood as follows. Let us consider how $\omega_a$ varies with $m_a$ in the two cases; this is shown in the left panel Fig.~\ref{fig:oma_vs_ma}. As above, we take $2\times 10^{-3}$ as a benchmark for the upper limit of hot relics allowed by Planck data. We do not need this value to be too precise for the argument that we are going to make. This threshold is reached for $m_a=0.75\,\eV$ ($\TRH = 100 \,\MeV$) or $m_a=0.45\,\eV$ ($\TRH = \infty$). Thus the observational limit is indeed violated sooner in the standard case with respect to $\TRH = 100 \,\MeV$. The upper limits on $m_a$ are however also determined by the shape of the posterior curves, which is quite different in the two cases, as it can be seen in the last two panels of Fig.~\ref{fig:post1D_ma}. It can be seen in Fig.~\ref{fig:oma_vs_ma} that, for  $\TRH = 100 \,\MeV$, the axion density is strongly suppressed  and basically negligible for $m_a<0.7\,\eV$. It then grows very quickly as the mass grows beyond this value, immediately overshooting the observational bound. The likelihood as a function of $m_a$ will thus be constant and equal to its value in $m_a=0$ up to $m_a=0.7\,\eV$, and then suddenly goes to zero for $m_a\gtrsim 0.7\,\eV$. This reflects in the shape of the posterior. In the standard case, the axion density is increasing more smoothly as a function of $m_a$. Thus the likelihood peaks at $m_a = 0$ but starts decreasing as soon as we move away from this value. There is however a relatively long tail related to the fact that axions become colder as the mass increases, and this somehow compensates for the larger density. This explains the different shape of the posteriors between $\TRH = 100 \,\MeV$ and $\TRH = \infty$.
It should be clear at this point that the region in $m_a$ encompassing the $95\%$ of the total posterior volume (the 95\% Bayesian credible interval) can indeed be larger for $\TRH = \infty$ than for $\TRH = 100 \,\MeV$. For the sake of clarity, we show the two posteriors together in the right panel of Fig.~\ref{fig:post1D_ma}, normalized so that they both integrate to unity.

Let us now comment on the bounds that we obtain from the extended datasets including also information from CMB lensing or BAO. 
Both additional datasets are particularly useful to constrain the abundance of light relics, either through their effect on structure formation (lensing) or by limiting possible changes to the expansion history (BAO). 
One might thus expect to find tighter constraints on the axion mass in at least some of the scenarios considered here, namely those in which the phenomenology of thermal axions is essentially that of a hot or warm particle. 
By looking at the values reported in Table~\ref{tab:ma_bounds} for low-reheating scenarios, we see that no significant improvement is observed for the \TPL\ dataset, with the exception of the $\TRH=20\,\MeV$ case; even in that case, it is quite marginal ($\sim 7\% $). 
For the \TPB\ dataset, the bounds improve in low-reheating scenarios with $20\,\MeV \le \TRH \le 50\,\MeV$, but also in this case the improvement is at most $\sim 10\%$ with respect to the value for \TPL.
The fact that there is no improvement for $\TRH=10\,\MeV$ can be easily understood from the fact, noted above, that in this case axions behave as cold particles and can make all the DM. 
Thus the relevant observational bound is the upper limit on the DM density that is not significantly improved by the inclusion of lensing or BAO data.~\footnote{Note that this statement holds for $\omega_a$ since this is bound to be $\le \omega_{cdm}$. The Planck determination of $\omega_{cdm}$ itself is indeed made more precise by the inclusion of BAO data.} For what concerns larger reheating temperatures, it is true that in these cases the axion phenomenology is that of a warm/hot particle; however, the axion density is a very steep function of mass around in the region around the observational upper limits, so the effects of tighter constraints on $\omega_a$ (the quantity more directly probed by observations) are strongly diluted when translated in terms of $m_a$. 
The steepness of the $(\omega_a,\,m_a)$ relation increase with the reheating temperature (see Fig.~\ref{fig:abund}), and this explains why the improvements become increasingly smaller and eventually vanish with larger $\TRH$. 
On the contrary, inclusion of lensing and BAO data significantly improves the constraints on the axion mass in the standard scenario: we obtain $m_a < 0.84,\, 0.64,\,0.26\,\eV$ for \TP, \TPL\ and \TPB, respectively. This improvements reflect the fact that, for the standard thermal history, the axion density varies relatively smoothly with their mass. Let us finally note that the observed dependence of the constraints from the dataset used is in agreement with the fact that the density of massive light relics is better constrained by adding current BAO information, rather than the measurements of the reconstructed lensing potential, to the Planck observations of temperature and polarization anisotropies.

We remark that our mass bounds in low-reheating temperature scenarios represent a significant improvement with respect to the seminal paper GSK08~\cite{Grin:2007yg}. For the largest value of the reheating temperature considered here, $\TRH \lesssim 100\,\MeV$,
GSK08 find $m_a \lesssim 2$~eV at 95\% C.L., as it can be inferred from a visual inspection of their Fig. 1. Our bound $m_a \lesssim 0.7\,\eV$ thus represents a factor three improvement with respect to that result. However, the most striking improvement is found for low values of the reheating temperature.
Indeed, in GSK08 it was found that the LSS/CMB bounds obtained from WMAP1/SDSS data, and based on free-streaming arguments, are completely lifted for $\TRH \lesssim 35\,\MeV$. Below that value of the reheating temperature,
bounds were obtained from the looser requirement that the axion density does not exceed the total dark matter density, as also discussed here in Sec. \ref{sec:axions_cosmo}. Given the precision of the data used in our analysis, we find that the lowest reheating temperature for which free streaming plays a role in determining constraints is lowered, lying somewhere in the $10-20\,\MeV$ range, as discussed in the previous sections. In GSK08, it is found that the free-streaming argument excludes masses larger than $20\,\eV$ for $\TRH \simeq 35 \,\MeV$, while the total density constraint yields $m_a < 200\,\eV$ for $\TRH \simeq 20\,\MeV$. For comparison, in the present analysis we have found that $m_a \lesssim 5\,\eV$ for $\TRH \simeq 35 \,\MeV$, and $m_a \lesssim 20\,\eV$ for $\TRH \simeq 20 \,\MeV$. We stress again that, in our case, free streaming still plays a role in determining the constrain at $\TRH \simeq 20 \,\MeV$.

We comment that for masses  $m_a > 18$ eV the axion lifetime with respect to radiative decay would be shorter than the  lifetime of the Universe~\cite{Cadamuro:2010cz}, unless the axion-photon coupling is suppressed~\cite{Kaplan:1985dv}. Therefore, our analysis might not apply in this regime. Note that axion decay might lead to other cosmological signatures, like e.g. changes in the ionization history of the Universe in the case of radiative decays~\cite{Cadamuro:2011fd}. 
Radiative decays of axions with masses of few eV might also affect the extra-galactic background light~\cite{Gong:2015hke,Kohri:2017oqn,Kalashev:2018bra,Caputo:2020msf}.
Given the bounds shown in Table~\ref{tab:ma_bounds}, this means that for $\TRH < 30$~MeV the mass of axions stable on cosmological timescales is essentially unconstrained by the data considered in this analysis.

As mentioned above, we also note that it was recently suggested~\cite{DiLuzio:2021vjd} that the chiral expansion effective field theory, upon which the computation of the axion-pion thermalization rate is based, might fail at temperatures larger than $T_\chi \simeq 62\,\MeV$. 
This would make the computation of the relic abundance of axions unreliable for the smaller masses that yield $T_D > T_\chi$. For the standard thermal history, this corresponds to $m_a \lesssim 1.1\,\eV$. We have computed the mass $m_{a,\chi}$ that yields $T_D = T_\chi$ for the reheating temperatures in consideration; these are shown in Tab.~\ref{tab:ma_chi}. We see that the allowed regions that we find for $\TRH \ge 50\,\MeV$ from the most constraining dataset, \TPB, fall completely within the region in which the validity of the chiral expansion has been questioned.
 
 \begin{table}[tbp]
\centering
\begin{tabular}{|c|c|}
\hline
\quad$\TRH \, [\MeV]$ & $m_{a,\chi} \,[\eV]$ \\
\hline
$10$& $12$\\
$ 20$& $6.0$\\
$ 30$& $3.9$\\
$ 40$& $2.9$\\
$ 50$& $2.3$\\
$ 80$& $1.1$\\
$ 100$&$1.1$ \\
$ \infty $& $1.1$ \\
\hline
\end{tabular}
\caption{Axion mass $m_{a,\chi}$ corresponding to a decoupling temperature $T_D$ equal to $T_\chi = 62\,\MeV$, for different values of the reheating temperature $\TRH$. The standard computation of the axion thermalization rate has been suggested to be unreliable at $T>T_\chi$ ~\cite{DiLuzio:2021vjd}. Axion with masses $m_a < m_{a,\chi}$ would decouple, according to the standard calculation, at $T_D > T_\chi$.
\label{tab:ma_chi}} 
\end{table}

\begin{figure}[t!]
\begin{center}
	\vspace{0.cm}
	\includegraphics[width=0.9\textwidth]{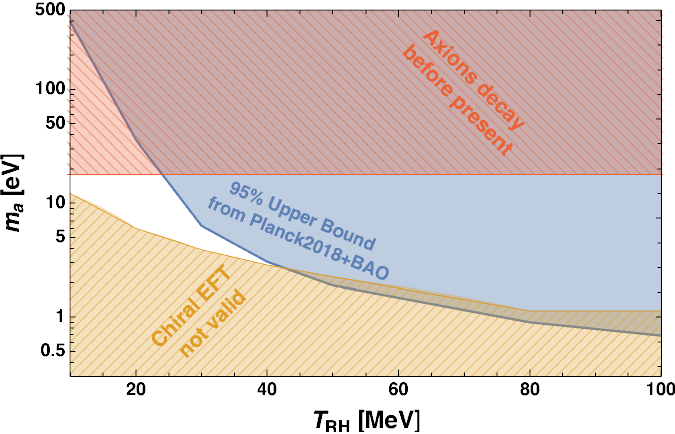}
		\caption{{Upper bounds on the axion mass as a function of the reheating temperature (solid blue line). The shaded blue area is excluded by our analysis, that neglects axion decays and assumes validity of the chiral effective field theory. The hatched areas mark the regions where i) the axion lifetime is larger than the present age of the Universe (red) (unless the axion photon coupling is suppressed), or ii) the computation of the axion-pion decay rate might fail (yellow, see text for discussion), and as such our analysis might not apply.}
	\label{fig:ma_bounds}}
\end{center}
\end{figure}

We summarize our findings in Fig.~\ref{fig:ma_bounds}, where we show the bounds on the axion mass from Planck temperature and polarization measurements in combination with BAO data, as a function of the reheating temperature. Even taking into account the finite lifetime of high-mass axions and the possible failure of chiral effective field theory in the small mass regime, our analysis shows that masses in the $3 - 20\,\eV$ range are allowed for sufficiently low values ($\TRH\le 40 \,\MeV$) of the reheating temperature. For values of the reheating temperature larger than $50\,\MeV$, instead, we find that all values of the mass for which the decoupling happens within the range of validity of the chiral EFT suggested by Ref.~\cite{DiLuzio:2021vjd} are excluded by the data.

\section{Conclusions}
We have re-examined cosmological bounds on thermal axions based on the 2018 CMB temperature anisotropy measurement provided by the Planck mission, as well as other types of cosmological observations. 
In the case of standard cosmological scenario we find and axion mass bound $m_a <0.259$ eV, comparable with the one recently placed in Ref.~\cite{Giare:2020vzo} using a similar dataset.
Our main goal has been to analyze how this bound would change in cosmological models with a low-reheating temperature $T_{\rm RH}$. We find that for $T_{\rm RH} \lesssim 80$ MeV the bound would relax to $m_a \lesssim {\mathcal O} (1)$ eV, becoming looser than $m_a \lesssim 10$ eV for $T_{\rm RH} \lesssim 30$ MeV. We remark that these reheating temperatures are still sufficiently high to leave unperturbed the neutrino thermalization~\cite{deSalas:2015glj,Hasegawa:2019jsa}. 
Therefore, axions would represent the only thermal relics sensitive to a range of reheating temperatures greater than the one probed by neutrinos ($T_{\rm RH} \sim 5$ MeV). 

Our result together with the recent one presented in~\cite{DiLuzio:2021vjd} questions the range of  validity of the cosmological mass bound on axions.
In~\cite{DiLuzio:2021vjd} it has been shown that for $m_a \lesssim 1$~eV the bound is not reliable since it is obtained by extrapolating the chiral expansion in a region where the effective field theory breaks down. Furthermore, we have shown that the bound is model dependent and would be significantly  relaxed in non-standard cosmologies, in a region in which other cosmological observables would not be affected.
Therefore, multi-eV axions do not seem necessarily excluded by cosmology alone. 
This finding  would motivate the necessity to come back to the astrophysical arguments in order to assess the robustness of the constrains for multi-eV axions. In particular, recent analyses seem to disfavor axions in this mass range from the supernova cooling~\cite{Carenza:2019pxu}. Our results would motivate a dedicated supernova simulation including multi-eV axions to definitely clarify this issue.

We remark that cosmological bounds on thermal axions are important in relation to direct search of axions at helioscope experiments~\cite{Arik:2013nya, Armengaud:2019uso}. 
In particular, the CERN Axion Solar Telescope (CAST) in their search for solar axions in the phase with  ${}^{3}\mathrm{He}$ buffer gas, has reached a  mass  $ m_a <1.17$ eV \cite{Arik:2013nya}.
However, CAST cannot detect larger masses  due to the loss of coherence of axion-photon conversions in the magnetic field of the detector. 
Also the future axion helioscope IAXO is not expected to reach a larger mass range. In this sense, the axion mass bound in the standard cosmological case is nicely complementary to the sensitivity of helioscopes search. Notably, it would allow to cover all the axion mass region around the eV without any gap. 
Our result shows that the cosmological bound is model-dependent and can be significantly loosened in presence of low-reheating temperatures,  which leave undisturbed the neutrino sector. 
Therefore, it seems mandatory to assess  new experimental  strategies to probe multi-eV axions. In this regard, there are ideas for a new class of  helioscopes, like the proposed  AMELIE (An Axion Modulation hELIoscope Experiment), which could be sensitive to axions with masses from a few $\rm{meV}$ to several $\eV$, thanks to the use of a Time Projection Chamber~\cite{Galan:2015msa}. 
Studies for low mass WIMPs are already being carried out by the TREX-DM experiment~\cite{Irastorza:2015dcb,Castel:2018gcp}, which is taking data at the Canfranc Underground Laboratory (LSC)~\cite{Castel:2019ngt}. 
The project aims at demonstrating the feasibility to reach low backgrounds at low energy thresholds for dark matter searches, which require similar detection conditions as for axions.
Furthermore, axions with mass $m_a \lesssim 100 \; \eV$ could be detected with a dark matter detector like the Cryogenic Underground Observatory for Rare Events (CUORE), which exploits the inverse Bragg-Primakoff effect to detect solar axions~\cite{Paschos:1993yf,Avignone:1999tv,Li:2015tsa}. 
It is also intriguing to mention that multi-eV would lead to an enhancement in the solar axion flux via resonant production in the solar magnetic field~\cite{Guarini:2020hps}.
Furthermore, axions with masses between 2-3~eV decaying into photons might explain the measured excess in the infrared photon background~\cite{Gong:2015hke,Kohri:2017oqn,Kalashev:2018bra,Caputo:2020msf}.
From a model-building perspective, there exist ``astrophobic'' models in which one would relax the axion mass bounds above eV~\cite{DiLuzio:2017ogq}.  
Furthermore, models have been conceived to accomodate together both hadronic axions and sterile neutrinos~\cite{Salvio:2015cja,Salvio:2021puw}. 
In this context, it has been recently proposed that low-reheating scenarios would reconcile eV-sterile neutrinos with cosmological observations~\cite{Hasegawa:2020ctq} (see also \cite{Gelmini:2004ah,Gelmini:2008fq,Benso:2019jog} for a study of the impact of low-reheating scenarios on heavy sterile neutrinos).
So if hints of eV-sterile neutrinos would be confirmed, they would strengthen the case for low-reheating scenarios and for multi-eV axions.

New detection possibilities need to be put on the agenda of the experimental strategies for axions, in order to explore  the multi-eV mass region where surprises might emerge. 
Indeed, if multi-eV axion should be found by one of these experiments, this would be unavoidably in tension with standard cosmological bound. 
Therefore, such a discovery would be not only revolutionary for particle physics, but would also produce a radical change in our description of the cosmological evolution of the Universe.

\section*{Acknowlegments}

We warmly thank Maurizio Giannotti, Steen Hannestad and Georg Raffelt for useful comments on our manuscript, and Martina 
Gerbino for useful discussions.
The work of P.C. and 
A.M. is partially supported by the Italian Istituto Nazionale di Fisica Nucleare (INFN) through the ``Theoretical Astroparticle Physics'' project
and by the research grant number 2017W4HA7S
``NAT-NET: Neutrino and Astroparticle Theory Network'' under the program
PRIN 2017 funded by the Italian Ministero dell'Universit\`a e della
Ricerca (MUR). M.L. and F.F. acknowledge support from the COSMOS network (www.cosmosnet.it) through the ASI (Italian Space Agency) Grants 2016-24-H.0, 2016-24-H.1-2018 and 2019-9-HH.0. We acknowledge the use of computing facilities provided by the INFN theory group (I.S. InDark) at CINECA.

\section*{Appendix: Axion decoupling in low-temperature reheating scenario}

In this Appendix we provide more details on axion thermalization in the low-temperature reheating scenario, closely following the discussion given in~\cite{Grin:2007yg}.
The radiation-dominated era of the Universe might be originated by the decay of a scalar field, that dominated the Universe until the decay.  The process of decay of a scalar field and thermalization of the produced particles is called reheating. In the history of the Universe one cannot exclude that the reheating phenomenon happened more than once. From a phenomenological point of view, we are interested only in the last reheating before the primordial nucleosynthesis. The field responsible for the reheating is usually associated with the inflaton and the temperature at which reheating occurs is denoted with $T_{\rm RH}$. Observations cannot exclude reheating temperatures as low as 1~MeV, and these models are called Low-Temperature Reheating scenarios. We suppose an initial phase dominated by a non-relativistic scalar field $\phi$ (for simplicity we call it inflaton), whose decay produces radiation that dominates the second phase.
At the beginning of reheating the Hubble parameter is determined entirely by the energy density of the inflaton
\begin{equation}
H_{0}^{2}=\frac{8\pi}{3m_{P}^{2}}\rho_{0,\phi}\;;
\end{equation}
and the Boltzmann equations for the inflaton and radiation energy densities are
\begin{equation}
\begin{split}
\dot{\rho}_{\phi}+3H\rho_{\phi}&=-\Gamma_\phi \rho_{\phi}\;,\\
\dot{\rho}_{R}+4H\rho_{R}&=\Gamma_\phi \rho_{\phi}\;;\\
\end{split}
\end{equation}
assuming a constant inflaton decay rate $\Gamma_\phi$. The two energy densities scale as radiation and non-relativistic matter. Then $\rho_{\phi}\sim a^{-3}e^{-\Gamma t}$, where the decay factor accounts for the inflaton decay. At the beginning of the reheating this term can be neglected and the radiation has an energy density
\begin{equation}
\rho_{R}=\Gamma \rho_{0,\phi}t_{0}\frac{3}{5}\left[\left(\frac{a_{0}}{a}\right)^{3/2}-\left(\frac{a_{0}}{a}\right)^{4}\right]\;;
\end{equation}
where $a_{0}$ is the scale factor at the beginning of reheating and it is easy to see that the maximum radiation density is achieved for $a_{m}=a_{0}(8/3)^{2/5}$.
Much later than $a_{0}$, the radiation density is proportional to $a^{-3/2}$. The temperature $T_{\rm RH}$ at which reheating is completed  is given by 
\begin{equation}
H^{2}=\Gamma_\phi^{2}=\frac{8\pi}{3m_{P}^{2}}\frac{\pi^{2}}{30}g_{*,RH}T_{\rm RH}^{4}\;;
\end{equation}
where $g_{*,RH}=g_{*}(T_{\rm RH})$ is the effective number of thermal degrees of freedom. 

\begin{figure}[t!]
	\vspace{0.cm}
	\hspace{1.cm}
	\includegraphics[width=0.8\textwidth]{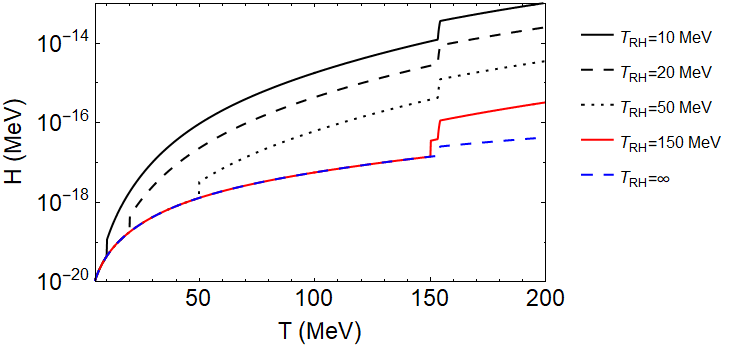}
	\caption{Evolution with temperature of the Hubble parameter, for different reheating temperatures $T_{\rm RH}$.}
	\label{fig:Hcomp}
\end{figure}

During the reheating phase, the temperature scales as
\begin{equation}
\begin{split}
T=\left(\frac{30}{\pi^{2}g_{*}(T)}\rho_{R}\right)^{1/4}=T_{m} \left(\frac{g_{*}(T_{m})}{g_{*}(T)}\right)^{1/4}\left(\frac{a_{m}}{a}\right)^{3/8}\;;
\end{split}
\end{equation}
where $T_{m}=T(a_{m})$. Thanks to the results above, one obtains the following Hubble parameter during reheating
\begin{equation}
H=\left[\frac{5}{9}\pi^{3}\frac{g_{*}(T)^{2}}{g_{*}(T_{\rm RH})}\right]^{1/2}\frac{T^{4}}{T_{\rm RH}^{2}m_{P}}\;;
\end{equation}
that can be compared with the standard scenario
\begin{equation}
H=\sqrt{\frac{4\pi^{3}}{45}g_{*}(T)}\frac{T^{2}}{m_{P}}\;. 
\end{equation}
Therefore, during reheating, the Universe expands faster than the standard scenario as can be seen from Fig.~\ref{fig:Hcomp},
where the jump in the behavior at $T=150$~MeV is due to the increase of the thermal degrees of freedom after the QCD phase transitions.

\begin{figure}[t!]
	\vspace{0.cm}
	\hspace{1.cm}
	\includegraphics[width=0.9\textwidth]{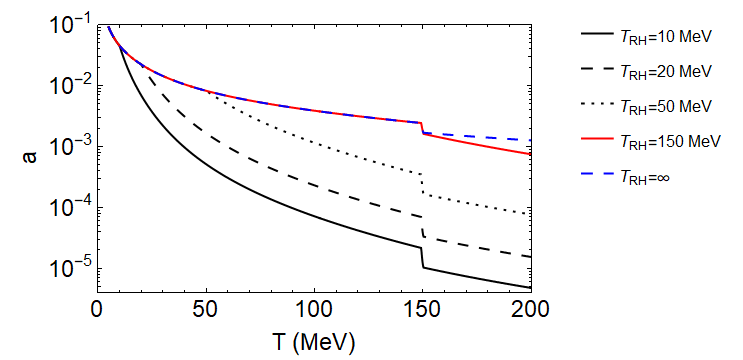}
	\caption{Scale factor for different reheating temperatures $T_{\rm RH}$.}
	\label{fig:scalef}
\end{figure}

The axion density at the decoupling  $n_{a,D}$ is diluted by the adiabatic Universe expansion as
\begin{equation}
\Omega_a=\frac{n_{a,D} m_{a}}{\rho_{c}}
\begin{cases}
\left(\frac{a_{D}}{a_{0}}\right)^{3}&{\rm for}\quad T_{D}< T_{\rm RH}\\
\left(\frac{a_{D}}{a_{\rm RH}}\right)^{3}\left(\frac{a_{\rm RH}}{a_{0}}\right)^{3}&{\rm for}\quad T_{D}\ge T_{\rm RH}\;;
\end{cases}
\label{eq:omega}
\end{equation}
where $\rho_{c}=1.05\times10^{4}h^{2}$~eV~${\rm cm}^{-3}$ is the critical density of the Universe.
The scale factor during the standard phase scales as
\begin{equation}
\begin{split}
\left(\frac{a_{\rm RH}}{a}\right)^{3}&=\left(\frac{T}{T_{\rm RH}}\right)^{3}\frac{g_{*,S}(T)}{g_{*,S}(T_{\rm RH})}\;;
\end{split}
\label{eq:scale1}
\end{equation}
and during the reheating phase as
\begin{equation}
\begin{split}
\left(\frac{a_{D}}{a_{\rm RH}}\right)^{3}&=\left(\frac{T_{\rm RH}}{T_{\rm D}}\right)^{8}\left[\frac{g_{*}(T_{RH})}{g_{*}(T_{\rm D})}\right]^{2}\;;
\end{split}
\label{eq:scale2}
\end{equation}
where the behaviour of these scale factors is given in Fig.~\ref{fig:scalef}.

Thus, from Eq.~(\ref{eq:omega}) the axion abundance is
\begin{equation}
\Omega_a h^{2}=\frac{m_{a}}{13~{\rm eV}}\frac{1}{g_{*,S}(T_{\rm D})}
\begin{cases}
1&{\rm for}\quad T_{D}< T_{\rm RH}\\
\left(\frac{T_{\rm RH}}{T_{\rm D}}\right)^{5}\left[\frac{g_{*}(T_{\rm RH})}{g_{*}(T_{\rm D})}\right]^{2}\left[\frac{g_{*,S}(T_{\rm D})}{g_{*,S}(T_{\rm RH})}\right]&{\rm for}\quad T_{D}\ge T_{\rm RH}\;;
\end{cases}
\end{equation}
the extra factor suppresses the relic abundance if $T_{\rm RH}< T_{\rm D}$.
In the standard scenario this term is equal to one. 
Note that this result does not depend on the initial value of the Hubble parameter $H_{0}$ or the maximum temperature $T_{m}$. An extra axion component in the radiation fluid, as long as it is ultrarelativisti, leads to an increase of the effective number of relativistic degrees of freedom
\begin{equation}
\begin{split}
N_{\rm eff}&=3+\left(\frac{\rho_{a}}{\rho_{\gamma}}\right)\left(\frac{8}{7}\right)\left(\frac{11}{4}\right)^{4/3}\;;
\end{split}
\end{equation}
where $\rho_{a}$ and $\rho_{\gamma}$ are the axion and photon energy densities, respectively. This deviation is quantified by 
\begin{equation}
\Delta N_{\rm eff}=\left(\frac{4}{7}\right)\left(\frac{11}{4}\right)^{4/3}\left(\frac{T_{\rm D}}{T}\right)^{4}
\begin{cases}
\left(\frac{T}{T_{\rm RH}}\right)^{4}\left[\frac{g_{*,S}(T)}{g_{*,S}(T_{\rm RH})}\right]^{4/3}&{\rm for}\quad T_{D}< T_{\rm RH}\\
\left(\frac{T_{RH}}{T_{D}}\right)^{32/3}\left[\frac{g_{*}(T_{RH})}{g_{*}(T_{\rm D})}\right]^{8/3}\left(\frac{T}{T_{RH}}\right)^{4}\left[\frac{g_{*,S}(T)}{g_{*,S}(T_{\rm RH})}\right]^{4/3}&{\rm for}\quad T_{D}\ge T_{\rm RH}\;;
\end{cases}
\end{equation}
where we used the scaling relations in Eqs.~\eqref{eq:scale1}-\eqref{eq:scale2}. Similar relations hold for the axion-to-photon temperature ratio, properly rescaling the axion decoupling temperature to account for the different cosmological phases
\begin{equation}
\mathcal{T}_{a}=\frac{T_{a}}{T}=\frac{T_{D}}{T}\frac{a_{D}}{a}=
\begin{cases}
\left[\frac{g_{*,S}(T)}{g_{*,S}(T_{D})}\right]^{1/3}&{\rm for}\quad T_{D}< T_{\rm RH}\\
\left[\frac{g_{*,S}(T)}{g_{*,S}(T_{RH})}\right]^{1/3}\left(\frac{T_{\rm RH}}{T_{\rm D}}\right)^{5/3}\left[\frac{g_{*}(T_{\rm RH})}{g_{*}(T_{\rm D})}\right]^{2/3}&{\rm for}\quad T_{D}\ge T_{\rm RH}\;.
\end{cases}
\end{equation}
Finally, the momentum-to-mass ratio is defined as
\begin{equation}
    \frac{\langle p_{a}\rangle}{m_{a}}=\frac{2.7 T_{a}}{m_{a}}=\frac{2.7 T}{m_{a}}\mathcal{T}_{a} \,\ ,
\end{equation}
where $T=T_{\rm today}(1+z)$ and we have calculated the average momentum by mediating over a Bose-Einstein distribution.

\bibliographystyle{bibi}
\bibliography{biblio.bib}

\providecommand{\href}[2]{#2}\begingroup\raggedright\begin{thebibliography}{100}

\bibitem{Lesgourgues:2006nd}
J.~Lesgourgues and S.~Pastor, \emph{{Massive neutrinos and cosmology}},
  \href{https://doi.org/10.1016/j.physrep.2006.04.001}{\emph{Phys. Rept.}
  {\bfseries 429} (2006) 307}
  [\href{https://arxiv.org/abs/astro-ph/0603494}{{\ttfamily
  astro-ph/0603494}}].

\bibitem{Lesgourgues:2014zoa}
J.~Lesgourgues and S.~Pastor, \emph{{Neutrino cosmology and Planck}},
  \href{https://doi.org/10.1088/1367-2630/16/6/065002}{\emph{New J. Phys.}
  {\bfseries 16} (2014) 065002}
  [\href{https://arxiv.org/abs/1404.1740}{{\ttfamily 1404.1740}}].

\bibitem{Archidiacono:2013fha}
M.~Archidiacono, E.~Giusarma, S.~Hannestad and O.~Mena, \emph{{Cosmic dark
  radiation and neutrinos}},
  \href{https://doi.org/10.1155/2013/191047}{\emph{Adv. High Energy Phys.}
  {\bfseries 2013} (2013) 191047}
  [\href{https://arxiv.org/abs/1307.0637}{{\ttfamily 1307.0637}}].

\bibitem{Capozzi:2017ipn}
F.~Capozzi, E.~Di~Valentino, E.~Lisi, A.~Marrone, A.~Melchiorri and A.~Palazzo,
  \emph{{Global constraints on absolute neutrino masses and their ordering}},
  \href{https://doi.org/10.1103/PhysRevD.95.096014}{\emph{Phys. Rev. D}
  {\bfseries 95} (2017) 096014}
  [\href{https://arxiv.org/abs/2003.08511}{{\ttfamily 2003.08511}}]. [Addendum:
  Phys.Rev.D 101, 116013 (2020)].

\bibitem{Lattanzi:2017ubx}
M.~Lattanzi and M.~Gerbino, \emph{{Status of neutrino properties and future
  prospects - Cosmological and astrophysical constraints}},
  \href{https://doi.org/10.3389/fphy.2017.00070}{\emph{Front. in Phys.}
  {\bfseries 5} (2018) 70} [\href{https://arxiv.org/abs/1712.07109}{{\ttfamily
  1712.07109}}].

\bibitem{RoyChoudhury:2019hls}
S.~Roy~Choudhury and S.~Hannestad, \emph{{Updated results on neutrino mass and
  mass hierarchy from cosmology with Planck 2018 likelihoods}},
  \href{https://doi.org/10.1088/1475-7516/2020/07/037}{\emph{JCAP} {\bfseries
  07} (2020) 037} [\href{https://arxiv.org/abs/1907.12598}{{\ttfamily
  1907.12598}}].

\bibitem{Hannestad:2003ye}
S.~Hannestad and G.~Raffelt, \emph{{Cosmological mass limits on neutrinos,
  axions, and other light particles}},
  \href{https://doi.org/10.1088/1475-7516/2004/04/008}{\emph{JCAP} {\bfseries
  04} (2004) 008} [\href{https://arxiv.org/abs/hep-ph/0312154}{{\ttfamily
  hep-ph/0312154}}].

\bibitem{Hannestad:2005df}
S.~Hannestad, A.~Mirizzi and G.~Raffelt, \emph{{New cosmological mass limit on
  thermal relic axions}},
  \href{https://doi.org/10.1088/1475-7516/2005/07/002}{\emph{JCAP} {\bfseries
  07} (2005) 002} [\href{https://arxiv.org/abs/hep-ph/0504059}{{\ttfamily
  hep-ph/0504059}}].

\bibitem{Melchiorri:2007cd}
A.~Melchiorri, O.~Mena and A.~Slosar, \emph{{An improved cosmological bound on
  the thermal axion mass}},
  \href{https://doi.org/10.1103/PhysRevD.76.041303}{\emph{Phys. Rev. D}
  {\bfseries 76} (2007) 041303}
  [\href{https://arxiv.org/abs/0705.2695}{{\ttfamily 0705.2695}}].

\bibitem{Hannestad:2007dd}
S.~Hannestad, A.~Mirizzi, G.~G. Raffelt and Y.~Y.~Y. Wong, \emph{{Cosmological
  constraints on neutrino plus axion hot dark matter}},
  \href{https://doi.org/10.1088/1475-7516/2007/08/015}{\emph{JCAP} {\bfseries
  08} (2007) 015} [\href{https://arxiv.org/abs/0706.4198}{{\ttfamily
  0706.4198}}].

\bibitem{Hannestad:2010yi}
S.~Hannestad, A.~Mirizzi, G.~G. Raffelt and Y.~Y.~Y. Wong, \emph{{Neutrino and
  axion hot dark matter bounds after WMAP-7}},
  \href{https://doi.org/10.1088/1475-7516/2010/08/001}{\emph{JCAP} {\bfseries
  08} (2010) 001} [\href{https://arxiv.org/abs/1004.0695}{{\ttfamily
  1004.0695}}].

\bibitem{Giusarma:2014zza}
E.~Giusarma, E.~Di~Valentino, M.~Lattanzi, A.~Melchiorri and O.~Mena,
  \emph{{Relic Neutrinos, thermal axions and cosmology in early 2014}},
  \href{https://doi.org/10.1103/PhysRevD.90.043507}{\emph{Phys. Rev. D}
  {\bfseries 90} (2014) 043507}
  [\href{https://arxiv.org/abs/1403.4852}{{\ttfamily 1403.4852}}].

\bibitem{DiValentino:2015wba}
E.~Di~Valentino, E.~Giusarma, M.~Lattanzi, O.~Mena, A.~Melchiorri and J.~Silk,
  \emph{{Cosmological Axion and neutrino mass constraints from Planck 2015
  temperature and polarization data}},
  \href{https://doi.org/10.1016/j.physletb.2015.11.025}{\emph{Phys. Lett. B}
  {\bfseries 752} (2016) 182}
  [\href{https://arxiv.org/abs/1507.08665}{{\ttfamily 1507.08665}}].

\bibitem{Archidiacono:2013cha}
M.~Archidiacono, S.~Hannestad, A.~Mirizzi, G.~Raffelt and Y.~Y.~Y. Wong,
  \emph{{Axion hot dark matter bounds after Planck}},
  \href{https://doi.org/10.1088/1475-7516/2013/10/020}{\emph{JCAP} {\bfseries
  10} (2013) 020} [\href{https://arxiv.org/abs/1307.0615}{{\ttfamily
  1307.0615}}].

\bibitem{Archidiacono:2015mda}
M.~Archidiacono, T.~Basse, J.~Hamann, S.~Hannestad, G.~Raffelt and Y.~Y.~Y.
  Wong, \emph{{Future cosmological sensitivity for hot dark matter axions}},
  \href{https://doi.org/10.1088/1475-7516/2015/05/050}{\emph{JCAP} {\bfseries
  05} (2015) 050} [\href{https://arxiv.org/abs/1502.03325}{{\ttfamily
  1502.03325}}].

\bibitem{DiValentino:2015zta}
E.~Di~Valentino, S.~Gariazzo, E.~Giusarma and O.~Mena, \emph{{Robustness of
  cosmological axion mass limits}},
  \href{https://doi.org/10.1103/PhysRevD.91.123505}{\emph{Phys. Rev. D}
  {\bfseries 91} (2015) 123505}
  [\href{https://arxiv.org/abs/1503.00911}{{\ttfamily 1503.00911}}].

\bibitem{Kim:1979if}
J.~E. Kim, \emph{{Weak Interaction Singlet and Strong CP Invariance}},
  \href{https://doi.org/10.1103/PhysRevLett.43.103}{\emph{Phys. Rev. Lett.}
  {\bfseries 43} (1979) 103}.

\bibitem{Shifman:1979if}
M.~A. Shifman, A.~I. Vainshtein and V.~I. Zakharov, \emph{{Can Confinement
  Ensure Natural CP Invariance of Strong Interactions?}},
  \href{https://doi.org/10.1016/0550-3213(80)90209-6}{\emph{Nucl. Phys. B}
  {\bfseries 166} (1980) 493}.

\bibitem{Giare:2020vzo}
W.~Giar\`e, E.~Di~Valentino, A.~Melchiorri and O.~Mena, \emph{{New cosmological
  bounds on hot relics: Axions $\&$ Neutrinos}},
  \href{https://arxiv.org/abs/2011.14704}{{\ttfamily 2011.14704}}.

\bibitem{Armengaud:2019uso}
{\scshape IAXO} Collaboration, E.~Armengaud et~al., \emph{{Physics potential of
  the International Axion Observatory (IAXO)}},
  \href{https://doi.org/10.1088/1475-7516/2019/06/047}{\emph{JCAP} {\bfseries
  06} (2019) 047} [\href{https://arxiv.org/abs/1904.09155}{{\ttfamily
  1904.09155}}].

\bibitem{Dine:1981rt}
M.~Dine, W.~Fischler and M.~Srednicki, \emph{{A Simple Solution to the Strong
  CP Problem with a Harmless Axion}},
  \href{https://doi.org/10.1016/0370-2693(81)90590-6}{\emph{Phys. Lett. B}
  {\bfseries 104} (1981) 199}.

\bibitem{Zhitnitsky:1980tq}
A.~R. Zhitnitsky, \emph{{On Possible Suppression of the Axion Hadron
  Interactions. (In Russian)}}, {\emph{Sov. J. Nucl. Phys.} {\bfseries 31}
  (1980) 260}.

\bibitem{Ferreira:2020bpb}
R.~Z. Ferreira, A.~Notari and F.~Rompineve,
  \emph{{Dine-Fischler-Srednicki-Zhitnitsky axion in the CMB}},
  \href{https://doi.org/10.1103/PhysRevD.103.063524}{\emph{Phys. Rev. D}
  {\bfseries 103} (2021) 063524}
  [\href{https://arxiv.org/abs/2012.06566}{{\ttfamily 2012.06566}}].

\bibitem{Baumann:2016wac}
D.~Baumann, D.~Green and B.~Wallisch, \emph{{New Target for Cosmic Axion
  Searches}}, \href{https://doi.org/10.1103/PhysRevLett.117.171301}{\emph{Phys.
  Rev. Lett.} {\bfseries 117} (2016) 171301}
  [\href{https://arxiv.org/abs/1604.08614}{{\ttfamily 1604.08614}}].

\bibitem{Abazajian:2016yjj}
{\scshape CMB-S4} Collaboration, K.~N. Abazajian et~al., \emph{{CMB-S4 Science
  Book, First Edition}},  \href{https://arxiv.org/abs/1610.02743}{{\ttfamily
  1610.02743}}.

\bibitem{Ferreira:2018vjj}
R.~Z. Ferreira and A.~Notari, \emph{{Observable Windows for the QCD Axion
  Through the Number of Relativistic Species}},
  \href{https://doi.org/10.1103/PhysRevLett.120.191301}{\emph{Phys. Rev. Lett.}
  {\bfseries 120} (2018) 191301}
  [\href{https://arxiv.org/abs/1801.06090}{{\ttfamily 1801.06090}}].

\bibitem{Irastorza:2018dyq}
I.~G. Irastorza and J.~Redondo, \emph{{New experimental approaches in the
  search for axion-like particles}},
  \href{https://doi.org/10.1016/j.ppnp.2018.05.003}{\emph{Prog. Part. Nucl.
  Phys.} {\bfseries 102} (2018) 89}
  [\href{https://arxiv.org/abs/1801.08127}{{\ttfamily 1801.08127}}].

\bibitem{Kawasaki:2000en}
M.~Kawasaki, K.~Kohri and N.~Sugiyama, \emph{{MeV scale reheating temperature
  and thermalization of neutrino background}},
  \href{https://doi.org/10.1103/PhysRevD.62.023506}{\emph{Phys. Rev. D}
  {\bfseries 62} (2000) 023506}
  [\href{https://arxiv.org/abs/astro-ph/0002127}{{\ttfamily
  astro-ph/0002127}}].

\bibitem{Hannestad:2004px}
S.~Hannestad, \emph{{What is the lowest possible reheating temperature?}},
  \href{https://doi.org/10.1103/PhysRevD.70.043506}{\emph{Phys. Rev. D}
  {\bfseries 70} (2004) 043506}
  [\href{https://arxiv.org/abs/astro-ph/0403291}{{\ttfamily
  astro-ph/0403291}}].

\bibitem{Ichikawa:2005vw}
K.~Ichikawa, M.~Kawasaki and F.~Takahashi, \emph{{The Oscillation effects on
  thermalization of the neutrinos in the Universe with low reheating
  temperature}}, \href{https://doi.org/10.1103/PhysRevD.72.043522}{\emph{Phys.
  Rev. D} {\bfseries 72} (2005) 043522}
  [\href{https://arxiv.org/abs/astro-ph/0505395}{{\ttfamily
  astro-ph/0505395}}].

\bibitem{deSalas:2015glj}
P.~F. de~Salas, M.~Lattanzi, G.~Mangano, G.~Miele, S.~Pastor and O.~Pisanti,
  \emph{{Bounds on very low reheating scenarios after Planck}},
  \href{https://doi.org/10.1103/PhysRevD.92.123534}{\emph{Phys. Rev. D}
  {\bfseries 92} (2015) 123534}
  [\href{https://arxiv.org/abs/1511.00672}{{\ttfamily 1511.00672}}].

\bibitem{Hasegawa:2019jsa}
T.~Hasegawa, N.~Hiroshima, K.~Kohri, R.~S.~L. Hansen, T.~Tram and S.~Hannestad,
  \emph{{MeV-scale reheating temperature and thermalization of oscillating
  neutrinos by radiative and hadronic decays of massive particles}},
  \href{https://doi.org/10.1088/1475-7516/2019/12/012}{\emph{JCAP} {\bfseries
  12} (2019) 012} [\href{https://arxiv.org/abs/1908.10189}{{\ttfamily
  1908.10189}}].

\bibitem{Hasegawa:2020ctq}
T.~Hasegawa, N.~Hiroshima, K.~Kohri, R.~S.~L. Hansen, T.~Tram and S.~Hannestad,
  \emph{{MeV-scale reheating temperature and cosmological production of light
  sterile neutrinos}},
  \href{https://doi.org/10.1088/1475-7516/2020/08/015}{\emph{JCAP} {\bfseries
  08} (2020) 015} [\href{https://arxiv.org/abs/2003.13302}{{\ttfamily
  2003.13302}}].

\bibitem{Chung:1998rq}
D.~J.~H. Chung, E.~W. Kolb and A.~Riotto, \emph{{Production of massive
  particles during reheating}},
  \href{https://doi.org/10.1103/PhysRevD.60.063504}{\emph{Phys. Rev. D}
  {\bfseries 60} (1999) 063504}
  [\href{https://arxiv.org/abs/hep-ph/9809453}{{\ttfamily hep-ph/9809453}}].

\bibitem{Kolb:2003ke}
E.~W. Kolb, A.~Notari and A.~Riotto, \emph{{On the reheating stage after
  inflation}}, \href{https://doi.org/10.1103/PhysRevD.68.123505}{\emph{Phys.
  Rev. D} {\bfseries 68} (2003) 123505}
  [\href{https://arxiv.org/abs/hep-ph/0307241}{{\ttfamily hep-ph/0307241}}].

\bibitem{Felder:1999wt}
G.~N. Felder, L.~Kofman and A.~D. Linde, \emph{{Gravitational particle
  production and the moduli problem}},
  \href{https://doi.org/10.1088/1126-6708/2000/02/027}{\emph{JHEP} {\bfseries
  02} (2000) 027} [\href{https://arxiv.org/abs/hep-ph/9909508}{{\ttfamily
  hep-ph/9909508}}].

\bibitem{Giudice:2000ex}
G.~F. Giudice, E.~W. Kolb and A.~Riotto, \emph{{Largest temperature of the
  radiation era and its cosmological implications}},
  \href{https://doi.org/10.1103/PhysRevD.64.023508}{\emph{Phys. Rev. D}
  {\bfseries 64} (2001) 023508}
  [\href{https://arxiv.org/abs/hep-ph/0005123}{{\ttfamily hep-ph/0005123}}].

\bibitem{Visinelli:2018wza}
L.~Visinelli and J.~Redondo, \emph{{Axion Miniclusters in Modified Cosmological
  Histories}}, \href{https://doi.org/10.1103/PhysRevD.101.023008}{\emph{Phys.
  Rev. D} {\bfseries 101} (2020) 023008}
  [\href{https://arxiv.org/abs/1808.01879}{{\ttfamily 1808.01879}}].

\bibitem{Blinov:2019rhb}
N.~Blinov, M.~J. Dolan, P.~Draper and J.~Kozaczuk, \emph{{Dark matter targets
  for axionlike particle searches}},
  \href{https://doi.org/10.1103/PhysRevD.100.015049}{\emph{Phys. Rev. D}
  {\bfseries 100} (2019) 015049}
  [\href{https://arxiv.org/abs/1905.06952}{{\ttfamily 1905.06952}}].

\bibitem{Blinov:2019jqc}
N.~Blinov, M.~J. Dolan and P.~Draper, \emph{{Imprints of the Early Universe on
  Axion Dark Matter Substructure}},
  \href{https://doi.org/10.1103/PhysRevD.101.035002}{\emph{Phys. Rev. D}
  {\bfseries 101} (2020) 035002}
  [\href{https://arxiv.org/abs/1911.07853}{{\ttfamily 1911.07853}}].

\bibitem{Ramberg:2019dgi}
N.~Ramberg and L.~Visinelli, \emph{{Probing the Early Universe with Axion
  Physics and Gravitational Waves}},
  \href{https://doi.org/10.1103/PhysRevD.99.123513}{\emph{Phys. Rev. D}
  {\bfseries 99} (2019) 123513}
  [\href{https://arxiv.org/abs/1904.05707}{{\ttfamily 1904.05707}}].

\bibitem{Grin:2007yg}
D.~Grin, T.~L. Smith and M.~Kamionkowski, \emph{{Axion constraints in
  non-standard thermal histories}},
  \href{https://doi.org/10.1103/PhysRevD.77.085020}{\emph{Phys. Rev. D}
  {\bfseries 77} (2008) 085020}
  [\href{https://arxiv.org/abs/0711.1352}{{\ttfamily 0711.1352}}].

\bibitem{Peccei:1977hh}
R.~D. Peccei and H.~R. Quinn, \emph{{CP Conservation in the Presence of
  Instantons}}, \href{https://doi.org/10.1103/PhysRevLett.38.1440}{\emph{Phys.
  Rev. Lett.} {\bfseries 38} (1977) 1440}.

\bibitem{Peccei:1977ur}
R.~D. Peccei and H.~R. Quinn, \emph{{Constraints Imposed by CP Conservation in
  the Presence of Instantons}},
  \href{https://doi.org/10.1103/PhysRevD.16.1791}{\emph{Phys. Rev. D}
  {\bfseries 16} (1977) 1791}.

\bibitem{Weinberg:1977ma}
S.~Weinberg, \emph{{A New Light Boson?}},
  \href{https://doi.org/10.1103/PhysRevLett.40.223}{\emph{Phys. Rev. Lett.}
  {\bfseries 40} (1978) 223}.

\bibitem{Wilczek:1977pj}
F.~Wilczek, \emph{{Problem of Strong $P$ and $T$ Invariance in the Presence of
  Instantons}}, \href{https://doi.org/10.1103/PhysRevLett.40.279}{\emph{Phys.
  Rev. Lett.} {\bfseries 40} (1978) 279}.

\bibitem{diCortona:2015ldu}
G.~Grilli~di Cortona, E.~Hardy, J.~Pardo~Vega and G.~Villadoro, \emph{{The QCD
  axion, precisely}},
  \href{https://doi.org/10.1007/JHEP01(2016)034}{\emph{JHEP} {\bfseries 01}
  (2016) 034} [\href{https://arxiv.org/abs/1511.02867}{{\ttfamily
  1511.02867}}].

\bibitem{Borsanyi:2016ksw}
S.~Borsanyi et~al., \emph{{Calculation of the axion mass based on
  high-temperature lattice quantum chromodynamics}},
  \href{https://doi.org/10.1038/nature20115}{\emph{Nature} {\bfseries 539}
  (2016) 69} [\href{https://arxiv.org/abs/1606.07494}{{\ttfamily 1606.07494}}].

\bibitem{Tanabashi:2018oca}
{\scshape Particle Data Group} Collaboration, M.~Tanabashi et~al.,
  \emph{{Review of Particle Physics}},
  \href{https://doi.org/10.1103/PhysRevD.98.030001}{\emph{Phys. Rev. D}
  {\bfseries 98} (2018) 030001}.

\bibitem{Giannotti:2017hny}
M.~Giannotti, I.~G. Irastorza, J.~Redondo, A.~Ringwald and K.~Saikawa,
  \emph{{Stellar Recipes for Axion Hunters}},
  \href{https://doi.org/10.1088/1475-7516/2017/10/010}{\emph{JCAP} {\bfseries
  10} (2017) 010} [\href{https://arxiv.org/abs/1708.02111}{{\ttfamily
  1708.02111}}].

\bibitem{Preskill:1982cy}
J.~Preskill, M.~B. Wise and F.~Wilczek, \emph{{Cosmology of the Invisible
  Axion}}, \href{https://doi.org/10.1016/0370-2693(83)90637-8}{\emph{Phys.
  Lett. B} {\bfseries 120} (1983) 127}.

\bibitem{Dine:1982ah}
M.~Dine and W.~Fischler, \emph{{The Not So Harmless Axion}},
  \href{https://doi.org/10.1016/0370-2693(83)90639-1}{\emph{Phys. Lett. B}
  {\bfseries 120} (1983) 137}.

\bibitem{Abbott:1982af}
L.~F. Abbott and P.~Sikivie, \emph{{A Cosmological Bound on the Invisible
  Axion}}, \href{https://doi.org/10.1016/0370-2693(83)90638-X}{\emph{Phys.
  Lett. B} {\bfseries 120} (1983) 133}.

\bibitem{Sikivie:2006ni}
P.~Sikivie, \emph{{Axion Cosmology}},
  \href{https://doi.org/10.1007/978-3-540-73518-2_2}{\emph{Lect. Notes Phys.}
  {\bfseries 741} (2008) 19}
  [\href{https://arxiv.org/abs/astro-ph/0610440}{{\ttfamily
  astro-ph/0610440}}].

\bibitem{Du:2018uak}
{\scshape ADMX} Collaboration, N.~Du et~al., \emph{{A Search for Invisible
  Axion Dark Matter with the Axion Dark Matter Experiment}},
  \href{https://doi.org/10.1103/PhysRevLett.120.151301}{\emph{Phys. Rev. Lett.}
  {\bfseries 120} (2018) 151301}
  [\href{https://arxiv.org/abs/1804.05750}{{\ttfamily 1804.05750}}].

\bibitem{Braine:2019fqb}
{\scshape ADMX} Collaboration, T.~Braine et~al., \emph{{Extended Search for the
  Invisible Axion with the Axion Dark Matter Experiment}},
  \href{https://doi.org/10.1103/PhysRevLett.124.101303}{\emph{Phys. Rev. Lett.}
  {\bfseries 124} (2020) 101303}
  [\href{https://arxiv.org/abs/1910.08638}{{\ttfamily 1910.08638}}].

\bibitem{TheMADMAXWorkingGroup:2016hpc}
{\scshape MADMAX Working Group} Collaboration, A.~Caldwell, G.~Dvali,
  B.~Majorovits, A.~Millar, G.~Raffelt, J.~Redondo, O.~Reimann, F.~Simon and
  F.~Steffen, \emph{{Dielectric Haloscopes: A New Way to Detect Axion Dark
  Matter}}, \href{https://doi.org/10.1103/PhysRevLett.118.091801}{\emph{Phys.
  Rev. Lett.} {\bfseries 118} (2017) 091801}
  [\href{https://arxiv.org/abs/1611.05865}{{\ttfamily 1611.05865}}].

\bibitem{Masso:2002np}
E.~Masso, F.~Rota and G.~Zsembinszki, \emph{{On axion thermalization in the
  early universe}},
  \href{https://doi.org/10.1103/PhysRevD.66.023004}{\emph{Phys. Rev. D}
  {\bfseries 66} (2002) 023004}
  [\href{https://arxiv.org/abs/hep-ph/0203221}{{\ttfamily hep-ph/0203221}}].

\bibitem{Graf:2010tv}
P.~Graf and F.~D. Steffen, \emph{{Thermal axion production in the primordial
  quark-gluon plasma}},
  \href{https://doi.org/10.1103/PhysRevD.83.075011}{\emph{Phys. Rev. D}
  {\bfseries 83} (2011) 075011}
  [\href{https://arxiv.org/abs/1008.4528}{{\ttfamily 1008.4528}}].

\bibitem{Salvio:2013iaa}
A.~Salvio, A.~Strumia and W.~Xue, \emph{{Thermal axion production}},
  \href{https://doi.org/10.1088/1475-7516/2014/01/011}{\emph{JCAP} {\bfseries
  01} (2014) 011} [\href{https://arxiv.org/abs/1310.6982}{{\ttfamily
  1310.6982}}].

\bibitem{Moroi:1998qs}
T.~Moroi and H.~Murayama, \emph{{Axionic hot dark matter in the hadronic axion
  window}}, \href{https://doi.org/10.1016/S0370-2693(98)01091-0}{\emph{Phys.
  Lett. B} {\bfseries 440} (1998) 69}
  [\href{https://arxiv.org/abs/hep-ph/9804291}{{\ttfamily hep-ph/9804291}}].

\bibitem{Chang:1993gm}
S.~Chang and K.~Choi, \emph{{Hadronic axion window and the big bang
  nucleosynthesis}},
  \href{https://doi.org/10.1016/0370-2693(93)90656-3}{\emph{Phys. Lett. B}
  {\bfseries 316} (1993) 51}
  [\href{https://arxiv.org/abs/hep-ph/9306216}{{\ttfamily hep-ph/9306216}}].

\bibitem{Raffelt:2006cw}
G.~G. Raffelt, \emph{{Astrophysical axion bounds}},
  \href{https://doi.org/10.1007/978-3-540-73518-2_3}{\emph{Lect. Notes Phys.}
  {\bfseries 741} (2008) 51}
  [\href{https://arxiv.org/abs/hep-ph/0611350}{{\ttfamily hep-ph/0611350}}].

\bibitem{Carenza:2019pxu}
P.~Carenza, T.~Fischer, M.~Giannotti, G.~Guo, G.~Mart\'\i{}nez-Pinedo and
  A.~Mirizzi, \emph{{Improved axion emissivity from a supernova via
  nucleon-nucleon bremsstrahlung}},
  \href{https://doi.org/10.1088/1475-7516/2019/10/016}{\emph{JCAP} {\bfseries
  10} (2019) 016} [\href{https://arxiv.org/abs/1906.11844}{{\ttfamily
  1906.11844}}]. [Erratum: JCAP 05, E01 (2020)].

\bibitem{Carenza:2020cis}
P.~Carenza, B.~Fore, M.~Giannotti, A.~Mirizzi and S.~Reddy, \emph{{Enhanced
  Supernova Axion Emission and its Implications}},
  \href{https://doi.org/10.1103/PhysRevLett.126.071102}{\emph{Phys. Rev. Lett.}
  {\bfseries 126} (2021) 071102}
  [\href{https://arxiv.org/abs/2010.02943}{{\ttfamily 2010.02943}}].

\bibitem{DiLuzio:2020wdo}
L.~Di~Luzio, M.~Giannotti, E.~Nardi and L.~Visinelli, \emph{{The landscape of
  QCD axion models}},
  \href{https://doi.org/10.1016/j.physrep.2020.06.002}{\emph{Phys. Rept.}
  {\bfseries 870} (2020) 1} [\href{https://arxiv.org/abs/2003.01100}{{\ttfamily
  2003.01100}}].

\bibitem{Capozzi:2020cbu}
F.~Capozzi and G.~Raffelt, \emph{{Axion and neutrino bounds improved with new
  calibrations of the tip of the red-giant branch using geometric distance
  determinations}},
  \href{https://doi.org/10.1103/PhysRevD.102.083007}{\emph{Phys. Rev. D}
  {\bfseries 102} (2020) 083007}
  [\href{https://arxiv.org/abs/2007.03694}{{\ttfamily 2007.03694}}].

\bibitem{Straniero:2020iyi}
O.~Straniero, C.~Pallanca, E.~Dalessandro, I.~Dominguez, F.~R. Ferraro,
  M.~Giannotti, A.~Mirizzi and L.~Piersanti, \emph{{The RGB tip of galactic
  globular clusters and the revision of the axion-electron coupling bound}},
  \href{https://doi.org/10.1051/0004-6361/202038775}{\emph{Astron. Astrophys.}
  {\bfseries 644} (2020) A166}
  [\href{https://arxiv.org/abs/2010.03833}{{\ttfamily 2010.03833}}].

\bibitem{Ayala:2014pea}
A.~Ayala, I.~Dom\'\i{}nguez, M.~Giannotti, A.~Mirizzi and O.~Straniero,
  \emph{{Revisiting the bound on axion-photon coupling from Globular
  Clusters}}, \href{https://doi.org/10.1103/PhysRevLett.113.191302}{\emph{Phys.
  Rev. Lett.} {\bfseries 113} (2014) 191302}
  [\href{https://arxiv.org/abs/1406.6053}{{\ttfamily 1406.6053}}].

\bibitem{Straniero:2015nvc}
O.~Straniero, A.~Ayala, M.~Giannotti, A.~Mirizzi and I.~Dominguez,
  \emph{{Axion-Photon Coupling: Astrophysical Constraints}},  in \emph{{11th
  Patras Workshop on Axions, WIMPs and WISPs}}, 2015,
  \href{https://doi.org/10.3204/DESY-PROC-2015-02/straniero_oscar}{DOI}.

\bibitem{Kaplan:1985dv}
D.~B. Kaplan, \emph{{Opening the Axion Window}},
  \href{https://doi.org/10.1016/0550-3213(85)90319-0}{\emph{Nucl. Phys. B}
  {\bfseries 260} (1985) 215}.

\bibitem{DiVecchia:1980yfw}
P.~Di~Vecchia and G.~Veneziano, \emph{{Chiral Dynamics in the Large n Limit}},
  \href{https://doi.org/10.1016/0550-3213(80)90370-3}{\emph{Nucl. Phys. B}
  {\bfseries 171} (1980) 253}.

\bibitem{Georgi:1986df}
H.~Georgi, D.~B. Kaplan and L.~Randall, \emph{{Manifesting the Invisible Axion
  at Low-energies}},
  \href{https://doi.org/10.1016/0370-2693(86)90688-X}{\emph{Phys. Lett. B}
  {\bfseries 169} (1986) 73}.

\bibitem{DiLuzio:2021vjd}
L.~Di~Luzio, G.~Martinelli and G.~Piazza, \emph{{Axion hot dark matter bound,
  reliably}},  \href{https://arxiv.org/abs/2101.10330}{{\ttfamily 2101.10330}}.

\bibitem{Husdal:2016haj}
L.~Husdal, \emph{{On Effective Degrees of Freedom in the Early Universe}},
  \href{https://doi.org/10.3390/galaxies4040078}{\emph{Galaxies} {\bfseries 4}
  (2016) 78} [\href{https://arxiv.org/abs/1609.04979}{{\ttfamily 1609.04979}}].

\bibitem{Aghanim:2018eyx}
{\scshape Planck} Collaboration, N.~Aghanim et~al., \emph{{Planck 2018 results.
  VI. Cosmological parameters}},
  \href{https://doi.org/10.1051/0004-6361/201833910}{\emph{Astron. Astrophys.}
  {\bfseries 641} (2020) A6}
  [\href{https://arxiv.org/abs/1807.06209}{{\ttfamily 1807.06209}}].

\bibitem{Kolb:1990vq}
E.~W. Kolb and M.~S. Turner, \emph{{The Early Universe}}, vol.~69. 1990.

\bibitem{Akrami:2018vks}
{\scshape Planck} Collaboration, N.~Aghanim et~al., \emph{{Planck 2018 results.
  I. Overview and the cosmological legacy of Planck}},
  \href{https://doi.org/10.1051/0004-6361/201833880}{\emph{Astron. Astrophys.}
  {\bfseries 641} (2020) A1}
  [\href{https://arxiv.org/abs/1807.06205}{{\ttfamily 1807.06205}}].

\bibitem{Alam:2016hwk}
{\scshape BOSS} Collaboration, S.~Alam et~al., \emph{{The clustering of
  galaxies in the completed SDSS-III Baryon Oscillation Spectroscopic Survey:
  cosmological analysis of the DR12 galaxy sample}},
  \href{https://doi.org/10.1093/mnras/stx721}{\emph{Mon. Not. Roy. Astron.
  Soc.} {\bfseries 470} (2017) 2617}
  [\href{https://arxiv.org/abs/1607.03155}{{\ttfamily 1607.03155}}].

\bibitem{Beutler:2011hx}
F.~Beutler, C.~Blake, M.~Colless, D.~H. Jones, L.~Staveley-Smith, L.~Campbell,
  Q.~Parker, W.~Saunders and F.~Watson, \emph{{The 6dF Galaxy Survey: Baryon
  Acoustic Oscillations and the Local Hubble Constant}},
  \href{https://doi.org/10.1111/j.1365-2966.2011.19250.x}{\emph{Mon. Not. Roy.
  Astron. Soc.} {\bfseries 416} (2011) 3017}
  [\href{https://arxiv.org/abs/1106.3366}{{\ttfamily 1106.3366}}].

\bibitem{Ross:2014qpa}
A.~J. Ross, L.~Samushia, C.~Howlett, W.~J. Percival, A.~Burden and M.~Manera,
  \emph{{The clustering of the SDSS DR7 main Galaxy sample \textendash{} I. A 4
  per cent distance measure at $z = 0.15$}},
  \href{https://doi.org/10.1093/mnras/stv154}{\emph{Mon. Not. Roy. Astron.
  Soc.} {\bfseries 449} (2015) 835}
  [\href{https://arxiv.org/abs/1409.3242}{{\ttfamily 1409.3242}}].

\bibitem{Aghanim:2019ame}
{\scshape Planck} Collaboration, N.~Aghanim et~al., \emph{{Planck 2018 results.
  V. CMB power spectra and likelihoods}},
  \href{https://doi.org/10.1051/0004-6361/201936386}{\emph{Astron. Astrophys.}
  {\bfseries 641} (2020) A5}
  [\href{https://arxiv.org/abs/1907.12875}{{\ttfamily 1907.12875}}].

\bibitem{Aghanim:2018oex}
{\scshape Planck} Collaboration, N.~Aghanim et~al., \emph{{Planck 2018 results.
  VIII. Gravitational lensing}},
  \href{https://doi.org/10.1051/0004-6361/201833886}{\emph{Astron. Astrophys.}
  {\bfseries 641} (2020) A8}
  [\href{https://arxiv.org/abs/1807.06210}{{\ttfamily 1807.06210}}].

\bibitem{Colombi:1995ze}
S.~Colombi, S.~Dodelson and L.~M. Widrow, \emph{{Large scale structure tests of
  warm dark matter}}, \href{https://doi.org/10.1086/176788}{\emph{Astrophys.
  J.} {\bfseries 458} (1996) 1}
  [\href{https://arxiv.org/abs/astro-ph/9505029}{{\ttfamily
  astro-ph/9505029}}].

\bibitem{Akita:2020szl}
K.~Akita and M.~Yamaguchi, \emph{{A precision calculation of relic neutrino
  decoupling}},
  \href{https://doi.org/10.1088/1475-7516/2020/08/012}{\emph{JCAP} {\bfseries
  08} (2020) 012} [\href{https://arxiv.org/abs/2005.07047}{{\ttfamily
  2005.07047}}].

\bibitem{Bennett:2020zkv}
J.~J. Bennett, G.~Buldgen, P.~F. De~Salas, M.~Drewes, S.~Gariazzo, S.~Pastor
  and Y.~Y.~Y. Wong, \emph{{Towards a precision calculation of $N_{\rm eff}$ in
  the Standard Model II: Neutrino decoupling in the presence of flavour
  oscillations and finite-temperature QED}},
  \href{https://doi.org/10.1088/1475-7516/2021/04/073}{\emph{JCAP} {\bfseries
  04} (2021) 073} [\href{https://arxiv.org/abs/2012.02726}{{\ttfamily
  2012.02726}}].

\bibitem{Cadamuro:2010cz}
D.~Cadamuro, S.~Hannestad, G.~Raffelt and J.~Redondo, \emph{{Cosmological
  bounds on sub-MeV mass axions}},
  \href{https://doi.org/10.1088/1475-7516/2011/02/003}{\emph{JCAP} {\bfseries
  02} (2011) 003} [\href{https://arxiv.org/abs/1011.3694}{{\ttfamily
  1011.3694}}].

\bibitem{Cadamuro:2011fd}
D.~Cadamuro and J.~Redondo, \emph{{Cosmological bounds on pseudo
  Nambu-Goldstone bosons}},
  \href{https://doi.org/10.1088/1475-7516/2012/02/032}{\emph{JCAP} {\bfseries
  02} (2012) 032} [\href{https://arxiv.org/abs/1110.2895}{{\ttfamily
  1110.2895}}].

\bibitem{Gong:2015hke}
Y.~Gong, A.~Cooray, K.~Mitchell-Wynne, X.~Chen, M.~Zemcov and J.~Smidt,
  \emph{{Axion decay and anisotropy of near-IR extragalactic background
  light}}, \href{https://doi.org/10.3847/0004-637X/825/2/104}{\emph{Astrophys.
  J.} {\bfseries 825} (2016) 104}
  [\href{https://arxiv.org/abs/1511.01577}{{\ttfamily 1511.01577}}].

\bibitem{Kohri:2017oqn}
K.~Kohri, T.~Moroi and K.~Nakayama, \emph{{Can decaying particle explain cosmic
  infrared background excess?}},
  \href{https://doi.org/10.1016/j.physletb.2017.07.026}{\emph{Phys. Lett. B}
  {\bfseries 772} (2017) 628}
  [\href{https://arxiv.org/abs/1706.04921}{{\ttfamily 1706.04921}}].

\bibitem{Kalashev:2018bra}
O.~E. Kalashev, A.~Kusenko and E.~Vitagliano, \emph{{Cosmic infrared background
  excess from axionlike particles and implications for multimessenger
  observations of blazars}},
  \href{https://doi.org/10.1103/PhysRevD.99.023002}{\emph{Phys. Rev. D}
  {\bfseries 99} (2019) 023002}
  [\href{https://arxiv.org/abs/1808.05613}{{\ttfamily 1808.05613}}].

\bibitem{Caputo:2020msf}
A.~Caputo, A.~Vittino, N.~Fornengo, M.~Regis and M.~Taoso, \emph{{Searching for
  axion-like particle decay in the near-infrared background: an updated
  analysis}}, \href{https://doi.org/10.1088/1475-7516/2021/05/046}{\emph{JCAP}
  {\bfseries 05} (2021) 046}
  [\href{https://arxiv.org/abs/2012.09179}{{\ttfamily 2012.09179}}].

\bibitem{Arik:2013nya}
{\scshape CAST} Collaboration, M.~Arik et~al., \emph{{Search for Solar Axions
  by the CERN Axion Solar Telescope with $^3$He Buffer Gas: Closing the Hot
  Dark Matter Gap}},
  \href{https://doi.org/10.1103/PhysRevLett.112.091302}{\emph{Phys. Rev. Lett.}
  {\bfseries 112} (2014) 091302}
  [\href{https://arxiv.org/abs/1307.1985}{{\ttfamily 1307.1985}}].

\bibitem{Galan:2015msa}
J.~Gal\'an et~al., \emph{{Exploring 0.1\textendash{}10 eV axions with a new
  helioscope concept}},
  \href{https://doi.org/10.1088/1475-7516/2015/12/012}{\emph{JCAP} {\bfseries
  12} (2015) 012} [\href{https://arxiv.org/abs/1508.03006}{{\ttfamily
  1508.03006}}].

\bibitem{Irastorza:2015dcb}
I.~G. Irastorza et~al., \emph{{Gaseous time projection chambers for rare event
  detection: Results from the T-REX project. I. Double beta decay}},
  \href{https://doi.org/10.1088/1475-7516/2016/01/033}{\emph{JCAP} {\bfseries
  01} (2016) 033} [\href{https://arxiv.org/abs/1512.07926}{{\ttfamily
  1512.07926}}].

\bibitem{Castel:2018gcp}
J.~Castel et~al., \emph{{Background assessment for the TREX Dark Matter
  experiment}},
  \href{https://doi.org/10.1140/epjc/s10052-019-7282-6}{\emph{Eur. Phys. J. C}
  {\bfseries 79} (2019) 782}
  [\href{https://arxiv.org/abs/1812.04519}{{\ttfamily 1812.04519}}].

\bibitem{Castel:2019ngt}
J.~Castel et~al., \emph{{The TREX-DM experiment at the Canfranc Underground
  Laboratory}}, \href{https://doi.org/10.1088/1742-6596/1468/1/012063}{\emph{J.
  Phys. Conf. Ser.} {\bfseries 1468} (2020) 012063}
  [\href{https://arxiv.org/abs/1910.13957}{{\ttfamily 1910.13957}}].

\bibitem{Paschos:1993yf}
E.~A. Paschos and K.~Zioutas, \emph{{A Proposal for solar axion detection via
  Bragg scattering}},
  \href{https://doi.org/10.1016/0370-2693(94)91233-5}{\emph{Phys. Lett. B}
  {\bfseries 323} (1994) 367}.

\bibitem{Avignone:1999tv}
{\scshape SOLAX} Collaboration, F.~T. Avignone et~al., \emph{{Solar axion
  experiments using coherent Primakoff conversion in single crystals}},
  \href{https://doi.org/10.1016/S0920-5632(98)00521-0}{\emph{Nucl. Phys. B
  Proc. Suppl.} {\bfseries 72} (1999) 176}.

\bibitem{Li:2015tsa}
D.~Li, R.~J. Creswick, F.~T. Avignone and Y.~Wang, \emph{{Theoretical Estimate
  of the Sensitivity of the CUORE Detector to Solar Axions}},
  \href{https://doi.org/10.1088/1475-7516/2015/10/065}{\emph{JCAP} {\bfseries
  10} (2015) 065} [\href{https://arxiv.org/abs/1507.00603}{{\ttfamily
  1507.00603}}].

\bibitem{Guarini:2020hps}
E.~Guarini, P.~Carenza, J.~Galan, M.~Giannotti and A.~Mirizzi,
  \emph{{Production of axionlike particles from photon conversions in
  large-scale solar magnetic fields}},
  \href{https://doi.org/10.1103/PhysRevD.102.123024}{\emph{Phys. Rev. D}
  {\bfseries 102} (2020) 123024}
  [\href{https://arxiv.org/abs/2010.06601}{{\ttfamily 2010.06601}}].

\bibitem{DiLuzio:2017ogq}
L.~Di~Luzio, F.~Mescia, E.~Nardi, P.~Panci and R.~Ziegler, \emph{{Astrophobic
  Axions}}, \href{https://doi.org/10.1103/PhysRevLett.120.261803}{\emph{Phys.
  Rev. Lett.} {\bfseries 120} (2018) 261803}
  [\href{https://arxiv.org/abs/1712.04940}{{\ttfamily 1712.04940}}].

\bibitem{Salvio:2015cja}
A.~Salvio, \emph{{A Simple Motivated Completion of the Standard Model below the
  Planck Scale: Axions and Right-Handed Neutrinos}},
  \href{https://doi.org/10.1016/j.physletb.2015.03.015}{\emph{Phys. Lett. B}
  {\bfseries 743} (2015) 428}
  [\href{https://arxiv.org/abs/1501.03781}{{\ttfamily 1501.03781}}].

\bibitem{Salvio:2021puw}
A.~Salvio and S.~Scollo, \emph{{Axion-Sterile-Neutrino Dark Matter}},
  \href{https://arxiv.org/abs/2104.01334}{{\ttfamily 2104.01334}}.

\bibitem{Gelmini:2004ah}
G.~Gelmini, S.~Palomares-Ruiz and S.~Pascoli, \emph{{Low reheating temperature
  and the visible sterile neutrino}},
  \href{https://doi.org/10.1103/PhysRevLett.93.081302}{\emph{Phys. Rev. Lett.}
  {\bfseries 93} (2004) 081302}
  [\href{https://arxiv.org/abs/astro-ph/0403323}{{\ttfamily
  astro-ph/0403323}}].

\bibitem{Gelmini:2008fq}
G.~Gelmini, E.~Osoba, S.~Palomares-Ruiz and S.~Pascoli, \emph{{MeV sterile
  neutrinos in low reheating temperature cosmological scenarios}},
  \href{https://doi.org/10.1088/1475-7516/2008/10/029}{\emph{JCAP} {\bfseries
  10} (2008) 029} [\href{https://arxiv.org/abs/0803.2735}{{\ttfamily
  0803.2735}}].

\bibitem{Benso:2019jog}
C.~Benso, V.~Brdar, M.~Lindner and W.~Rodejohann, \emph{{Prospects for Finding
  Sterile Neutrino Dark Matter at KATRIN}},
  \href{https://doi.org/10.1103/PhysRevD.100.115035}{\emph{Phys. Rev. D}
  {\bfseries 100} (2019) 115035}
  [\href{https://arxiv.org/abs/1911.00328}{{\ttfamily 1911.00328}}].

\end{thebibliography}\endgroup

\end{document}